\documentclass[aps,prl,twocolumn,superscriptaddress]{revtex4-2}
\usepackage{graphicx}

\graphicspath{./figure}

\begin{document}

\title{Structural Transition and Magnetic Anisotropy in $\alpha$-RuCl$_{3}$} 



\author{Subin~Kim}
\author{Ezekiel Horsley}
\affiliation{Department of Physics, University of Toronto, Toronto, Ontario, M5S 1A7, Canada}

\author{Jacob P. C. Ruff}
\affiliation{CHESS, Cornell University, Ithaca, New York 14853, USA}

\author{Beatriz D. Moreno}
\affiliation{Canadian Light Source Inc., 44 Innovation Boulevard, Saskatoon, SK S7N 2V3, Canada}

\author{Young-June~Kim}
\affiliation{Department of Physics, University of Toronto, Toronto, Ontario, M5S 1A7, Canada}
\email{youngjune.kim@utoronto.ca}


\date{\today}

\begin{abstract}

We report X-ray diffraction and magnetic susceptibility studies of the structural phase transition in $\alpha$-RuCl$_{3}$. By utilizing a single crystal sample with predominantly single twin domain, we show that $\alpha$-RuCl$_{3}$ goes from high-temperature C2/m structure to a rhombohedral structure with R$\bar{3}$ symmetry at low temperature. While the defining feature of the structural transition is changing the stacking direction from the monoclinic a-axis to the b-axis, bond-anisotropy disappears when the structural change occurs, indicating that the local $C_3$ symmetry is restored within the honeycomb layer. The symmetry change is corroborated by the vanishing magnetic anisotropy in the low-temperature structure. Our study demonstrates that magnetic interaction is extremely sensitive to structural details in $\alpha$-RuCl$_{3}$, which could explain the sample dependence found in this material.
\end{abstract}

\maketitle 



\begin{figure} [ht]
    \centering
    \includegraphics[width=0.5\textwidth]{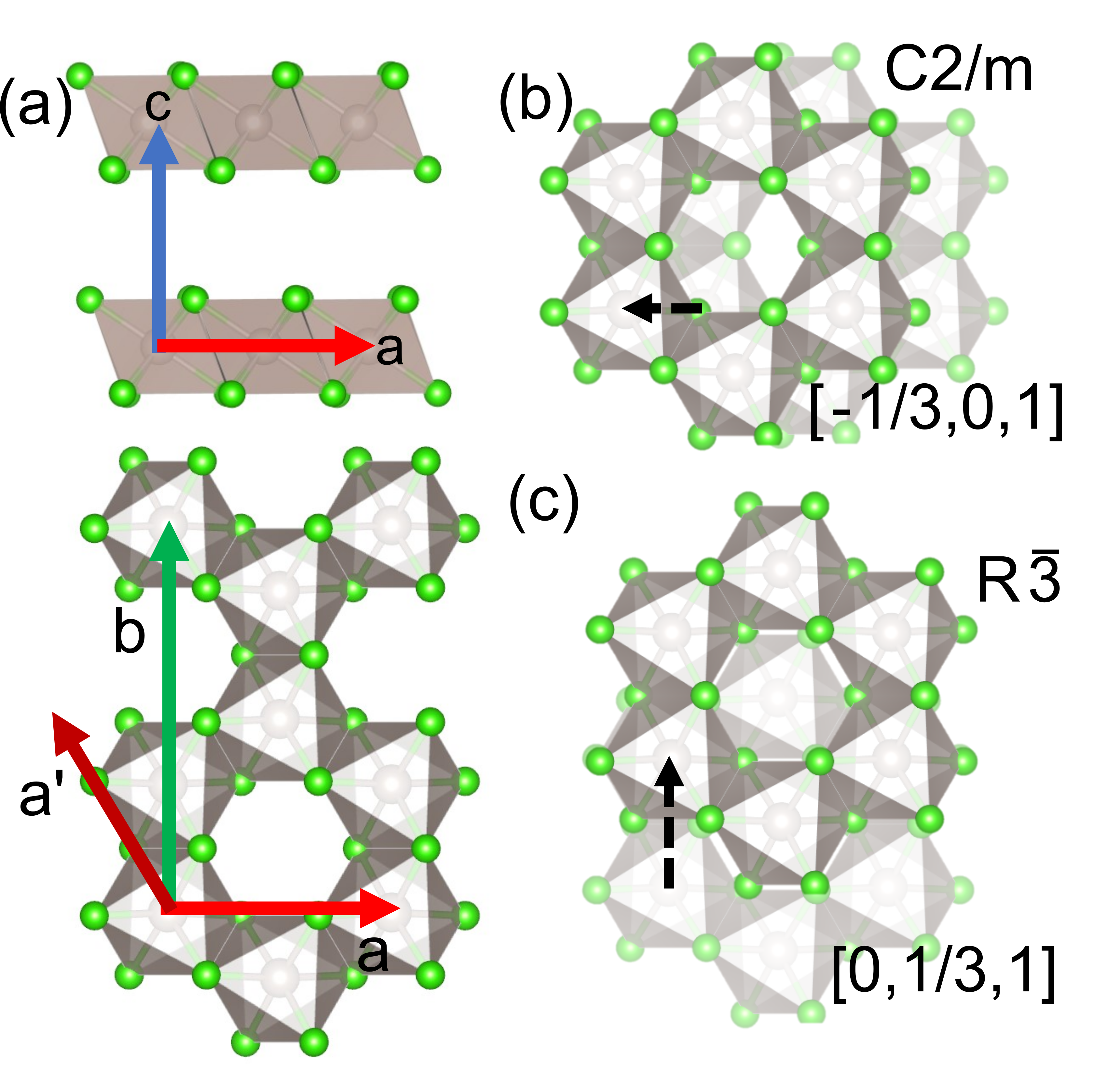}
    \caption{Crystal structure of $\alpha$-RuCl$_{3}$. (a) Orthorhombic axes used in this Letter. (b) Structure of $\alpha$-RuCl$_{3}$ above the structural transition temperature. Monoclinic C2/m structure has neighbouring honeycomb layers that are shifted along [-1/3,0,1]. (c) Structure of $\alpha$-RuCl$_{3}$ below the structural transition temperature. Rhombohedral R$\bar{3}$ structure has neighbouring honeycomb layers that are shifted along [0,1/3,1]. }
    \label{fig:structure}
\end{figure}

In recent years, $\alpha$-RuCl$_{3}$ has emerged as the prime candidate for realizing a Kitaev quantum spin liquid phase \cite{Plumb2014,sears15,banerjee16,Do2017,Banerjee2018,sandilands15,majumder15,kubota15,johnson15,cao16,leahy2017,baek2017,Sears2017,wolter2017,zheng2017,Kasahara2018,lampenkelley2018,modic2018NC,yokoi2021,bruin2022,Lefrancois2022,Winter2017,Takagi2019,Motome2020review,Kim2022}. In $\alpha$-RuCl$_{3}$, strong spin-orbit coupling (SOC) and the honeycomb network formed by edge-sharing RuCl$_6$ octahedra  (see Fig.~\ref{fig:structure}) provides a platform to realize a bond-dependent anisotropic interaction called Kitaev interaction ($K$), an essential ingredient for realizing Kitaev's honeycomb model \cite{kitaev2006,Winter2017,Takagi2019,Motome2020review,Hermanns2018}. Although $\alpha$-RuCl$_{3}$ magnetically orders below 7K, this order can be fully suppressed via applying magnetic field \cite{majumder15,kubota15,leahy2017,baek2017,Sears2017,wolter2017,zheng2017}.  The explosion of interest in this material was spurred by the discovery of a half-quantized thermal Hall effect in this phase, suggesting that this field-induced phase is a quantum spin liquid with Majorana fermions as heat carriers \cite{Kasahara2018}. However, subsequent experimental reports seem to suggest that the half-quantized thermal Hall effect is highly sample dependent \cite{czajka2021,bruin2022,Lefrancois2022}. 
Theoretical studies found that in addition to the Kitaev interaction, off-diagonal symmetric exchange interaction $\Gamma$ as well as isotropic Heisenberg interaction $J$ are important for describing the physics of $\alpha$-RuCl$_{3}$ \cite{Rau2014,winter2017NC,janssen2017,modic2018,modic2018NC,ran2017,riedl2019,koitzsch2020,Sears2020}. In addition, further neighbor interactions or additional off-diagonal terms due to trigonal distortion are often considered in the study of $\alpha$-RuCl$_{3}$ \cite{Lee2016,Kimchi2015,Winter2016,Winter2017,gordon2019,Sears2020}. Due to the complexity of the model, there is no consensus on the size (and sometimes even signs) of these interaction terms.

Another defining characteristic of $\alpha$-RuCl$_{3}$ is that it belongs to a family of magnetic van der Waals materials with an easily cleavable layered structure. While this opens up the exciting possibility of using $\alpha$-RuCl$_{3}$ in van der Waals heterostructures, it also means that this material is susceptible to the proliferation of stacking faults. It is now widely accepted that high-quality samples with a minimal number of stacking faults are in the monoclinic C2/m structure at room temperature \cite{johnson15,cao16,park2016}. These samples are characterized by a single magnetic transition around $T_N=7$~K, while samples with many stacking faults tend to show multiple transitions in the range of 10~K to 14~K \cite{kubota15,banerjee16,Kim2022}. It turns out that even the high-quality samples show small differences in $T_N$, ranging from 6.5~K to 8~K \cite{Kasahara2022,Zhang2023}. This additional sample variability is closely associated with the first-order structural phase transition around 150~K, which changes the stacking structure at low temperatures \cite{kubota15,park2016,Glamazda2017,Mu2022,Zhang2023}. The twinning of the low-temperature structure could introduce a large number of stacking faults even for a high-quality (at room temperature) sample, which also makes it difficult to determine the low-temperature crystal structure unambiguously \cite{Mu2022,Zhang2023}.

While one might question whether the stacking sequence matters for two-dimensional Kitaev physics in $\alpha$-RuCl$_{3}$, the experimentally observed sample dependence of the half-quantized thermal Hall effect suggests it does \cite{Kasahara2018,czajka2021,bruin2022,Lefrancois2022}. An interesting question is whether the structural difference implies a difference in the underlying magnetic Hamiltonian, which would be unaffected if the structural difference is strictly due to the stacking sequence of honeycomb layers. To answer this question, one should pay attention to the local symmetry that governs magnetic interactions, rather than the global structural symmetry, which is determined from stacking arrangements. 
	
In this Letter, we report our detailed investigation of structural and magnetic properties of a high-quality single crystal $\alpha$-RuCl$_{3}$ across the structural phase transition. This is made possible by studying a low-temperature-twin-free single crystal sample. We find that the low-temperature structure has a rhombohedral R$\bar{3}$ symmetry, arising from stacking of neighboring layers along the monoclinic b-direction. Crucially, this low-temperature structure recovers the $C_3$ rotational symmetry of the honeycomb plane, which is broken in the room-temperature C2/m structure. This symmetry change is corroborated by our bond-length data as well as magnetic susceptibility data. The implication is that the magnetic Hamiltonian of $\alpha$-RuCl$_{3}$ must have $C_3$ symmetry, although this symmetry might be fragile against structural stacking disorder, such as the coexistence of C2/m and R$\bar{3}$ stacking due to incomplete structural transformation.

{\em Experimental details:}
Single crystal $\alpha$-RuCl$_{3}$ crystals were grown using chemical vapor transport methods as described in Ref.~\cite{Kim2022}. Carefully selected crystals show a sharp single magnetic transition at $T_N$=7.2~K with a sharp mosaic width along $L$ of less than 0.1 degrees. Magnetic susceptibility was measured using Quantum Design Magnetic Property Measurement System (MPMS) and specific heat was measured using Quantum Design Physical Property Measurement System (PPMS). Single crystal X-ray diffraction measurements were carried out at the BXDS-IVU beamline at Canadian Light Source (CLS) with 10~keV X-ray energy and also using the Rigaku Smartlab diffractometer at the University of Toronto. The reciprocal space maps were obtained at the QM2 beamline at Cornell High Energy Synchrotron Source (CHESS) using 20~keV X-ray energy.

\noindent
{\em Structural Transition}: 
The crystal structure of $\alpha$-RuCl$_{3}$ is shown in Fig.~\ref{fig:structure}. We find it convenient to describe both monoclinic and rhombohedral structures using an orthorhombic coordinate system shown in Fig.~\ref{fig:structure}(a). The in-plane unit vectors $\vec{a}_{o}$ and $\vec{b}_{o}$ are two distinct high-symmetry vectors, same as in the C2/m structure, but $\vec{c}_o$ now refers to the vector perpendicular to the honeycomb plane with the length equal to the layer separation. Note that $\vec{c}_o$ is not a lattice translation vector in either structure and one should be careful when comparing $h,k,l$ in different structures. See Supplemental Material for reciprocal space comparisons. We will drop the subscript in the following discussions. The defining feature of the structural transition is the change in how the top layer is stacked against the bottom layer. As we will show below, the neighboring layer on top is shifted along the $\vec{a}$ direction in the high-temperature structure, while the shift direction changes to $\vec{b}_{o}$ below the structural transition temperature. The stacking sequence in the high-temperature C2/m structure can be restated as the lattice translation vector $-\frac{1}{3}\vec{a}+\vec{c}$ in the three-layer periodic structure. The stacking sequence in the R$\bar{3}$ structure is $\pm \frac{1}{3}\vec{b}+\vec{c}$. Note that there are two equivalent translation vectors in the R$\bar{3}$ structure. The C2/m and R$\bar{3}$ structures can be distinguished easily in a diffraction experiment. For the C2/m structure with $-\frac{1}{3}\vec{a}+\vec{c}$ translation, we expect Bragg peaks to occur at $(h,k,l+h/3)$ where $h,k,l$ are integers with even $h+k$. Now, for the lattice translation vector $\pm \frac{1}{3}\vec{b}+\vec{c}$, Bragg peaks will be found at $(h,k,l \mp k/3)$.

\begin{figure} [ht]
    \centering
    \includegraphics[width=0.5\textwidth]{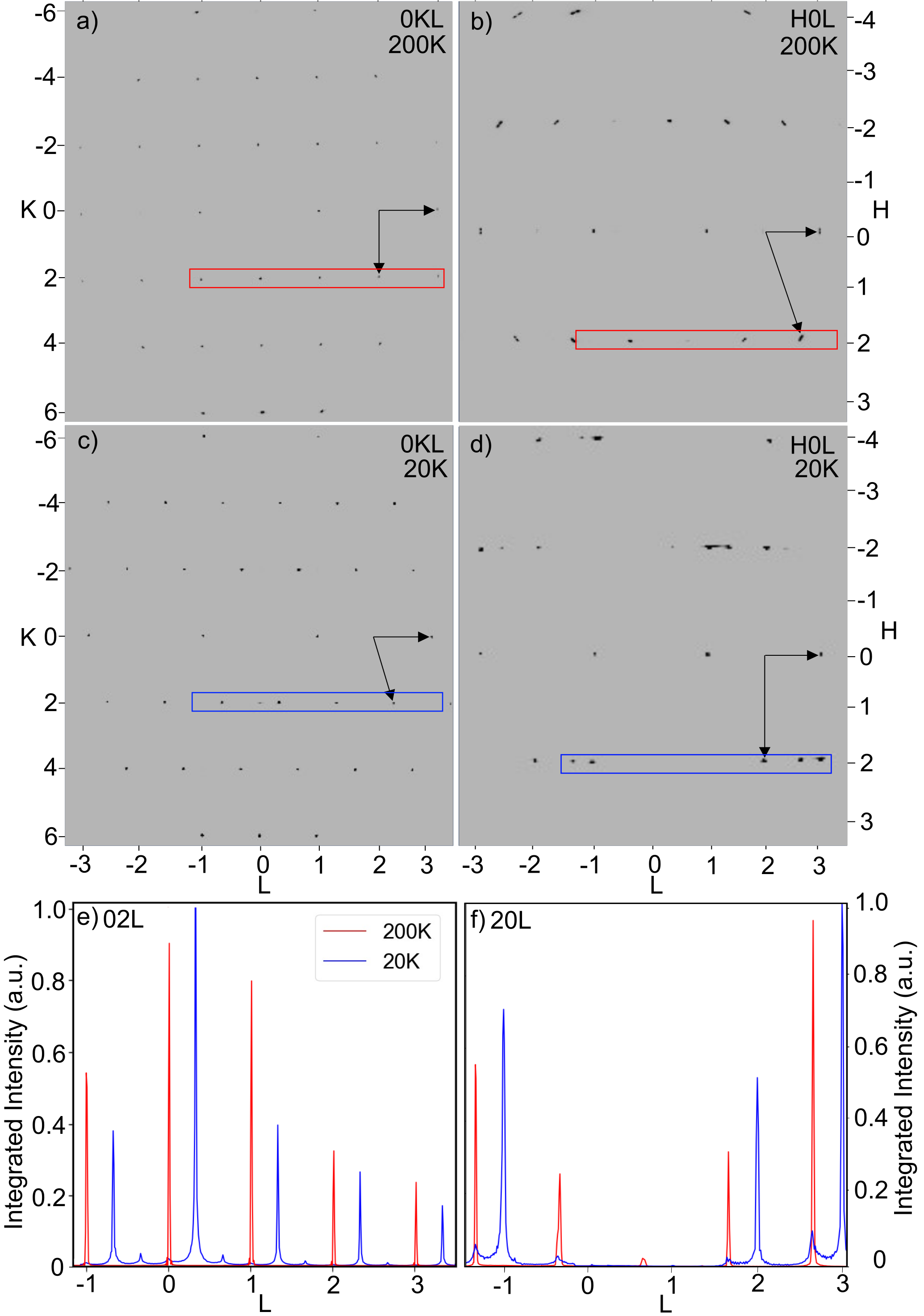}
    \caption{(a)(b) X-ray diffraction reciprocal space map of (0,K,L) plane  and (H,0,L) plane respectively above the structural transition of 200K. The peak positions can be well-explained using monoclinic structures which is shown in Fig.1(b). (c)(d) same reciprocal space map at 20K. The main peaks are well explained by the rhombohedral structure shown in Fig.1(c). (e)(f) shows L scan along (0,2,L) and (2,0,L) in the rectangular boxes shown in panels (a)-(d), demonstrating the change in Bragg peak positions across the structural transition.}
    \label{fig:xrd}
\end{figure}

Figure~\ref{fig:xrd} shows x-ray diffraction reciprocal space maps and line scans at two different temperatures above and well below the structural transition temperature. In Fig.~\ref{fig:xrd}(a)(b), the reciprocal space maps at 200K are shown. Clear sharp Bragg peaks with little diffuse scattering is observed, which confirms the high crystalline quality with minimum stacking faults at high temperature. As expected from the structure factor introduced above, Bragg peaks are observed at all integer $L$ values in the $(0,K,L)$ plane, while they are observed at non-integer $L$ values in the $(H,0,L)$ plane. Figure~\ref{fig:xrd}(c)(d) shows reciprocal space maps at 20~K, showing the shift of the Bragg peak position below the transition temperature. To see this clearly, in Fig.~\ref{fig:xrd}(e)(f), the intensity in the rectangular boxes is plotted as a function of $L$ for both $(0,2,L)$ and $(2,0,L)$ directions. The $(0,2,l)$ Bragg peaks shift to $(0,2,l-2/3)$ -- equivalently $(0,2,l+1/3$) -- and the $(2,0,l+2/3)$ Bragg peaks shift to $(2,0,l)$, respectively. This change in the Bragg peak positions is precisely what is expected from a transition from the monoclinic C2/m structure to the rhombohedral R$\bar{3}$ structure. In addition, this shift can be explained with a single translation vector $\frac{1}{3}\vec{b}+\vec{c}$, associated with only one type of twin domain. In fact, one can observe small peaks at $(0,2,l+2/3)$ in Fig.~\ref{fig:xrd}(e), which is due to contributions from the minority twin domain. We estimate more than 95\% of the crystal is in the majority twin domain. On the other hand, a more even mixture of the two twin domains is found in many other crystals we examined, as is the data shown in the recent study by Zhang et al. \cite{Zhang2023}. We also note that our data rules out the $P3_1 12$ stacking, which would show both $l \pm 1/3$ peaks.
However, even for our sample, which consists of mostly single twin-domain, weak diffuse scattering develops at low temperatures, indicating the presence of stacking faults. 

The $C2/m$ or $R\bar{3}$ crystal symmetry does not necessarily mean that the honeycomb layer must have the same local symmetry. An ideal honeycomb layer has local $C_3$ symmetry with the rotation axis out of the plane. However, monoclinic stacking of the honeycomb layers in the C2/m structure breaks global three-fold rotational symmetry. On the other hand, the stacking shift along the b-axis of the R$\bar{3}$ structure (Fig.~\ref{fig:structure}(c)) preserves the $C_3$ symmetry. As shown below, we find evidence from our x-ray diffraction that the local symmetry also changes when the structural change occurs. That is, the structural transition is not just a stacking sequence change, but is accompanied by the in-plane bond-length change. 

\begin{figure} [ht]
    \centering
    \includegraphics[width=0.5\textwidth]{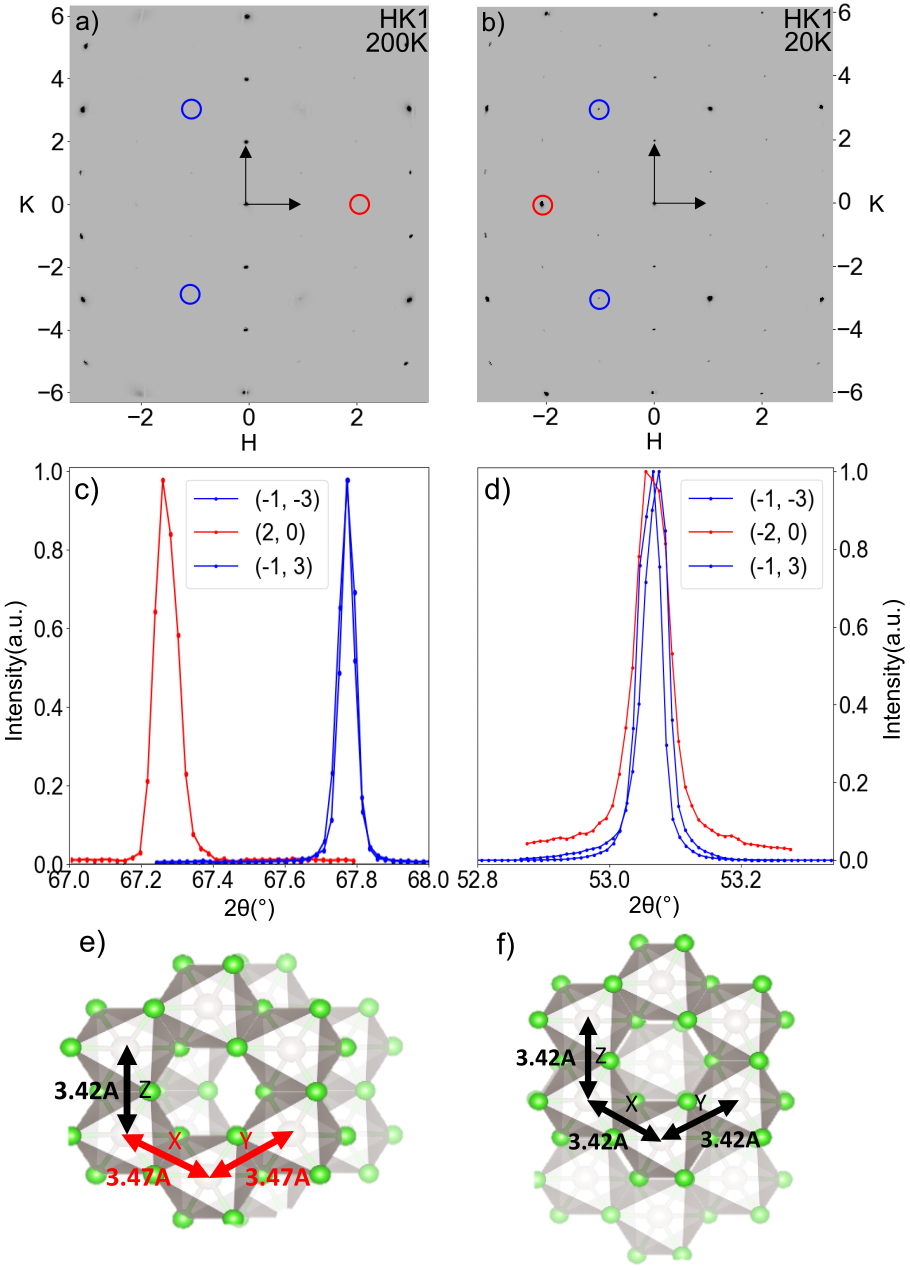}
    \caption{(a)(b) X-ray diffraction reciprocal space map of the $(H,K)$ plane with fixed $L=1$, obtained at 200~K and 20~K, respectively. (c)(d) High-resolution $2\theta$ scans of the peaks equivalent to the ones circled in panels (a) and (b), respectively. Note that $L=3.67$ and $L=5$ are chosen for panels (c) and (d), respectively, because of the shift of the Bragg peak discussed in Figure 2. A clear difference in the $2\theta$ values is observed at 200~K, which disappears at 20~K. (e)(f) The structure with an in-plane lattice parameter above and below the structural transition.}
    \label{fig:bond}
\end{figure}

In Fig.~\ref{fig:bond}(c,d), we compare $2\theta$ scans of three equivalent peaks at two temperatures. The $(-2,0,L)$, $(-1,-3,L)$, and $(-1,3,L)$ Bragg peaks with common $L$ would be symmetry equivalent in an ideal honeycomb structure. These peak positions are denoted with circles in the reciprocal space map shown in Fig.~\ref{fig:bond}(b). This is the case in the low-temperature structure with a good agreement in $2\theta$ values between the three peaks. In contrast, a clear difference in $2\theta$ is observed at high temperatures, indicating the presence of inequivalent bonds. Note that we need to use $(2,0,L)$ instead of $(-2,0,L)$ in the $C2/m$ structure to find equivalent peaks (see Supplemental Material for further explanation). We find that the bonds along the stacking direction in the monoclinic structure are elongated as shown in Fig.~\ref{fig:bond}. In other words, the bond-anisotropy present at high temperatures vanishes below the structural transition temperature, and the crystal recovers global as well as local $C_3$ symmetry at low temperatures. Our result, therefore, indicates that the structural distortion observed in the exfoliated monolayer sample~\cite{yang_magnetic_2023} is not present in a bulk crystal.

\begin{figure} [ht]
    \centering
    \includegraphics[width=0.5\textwidth]{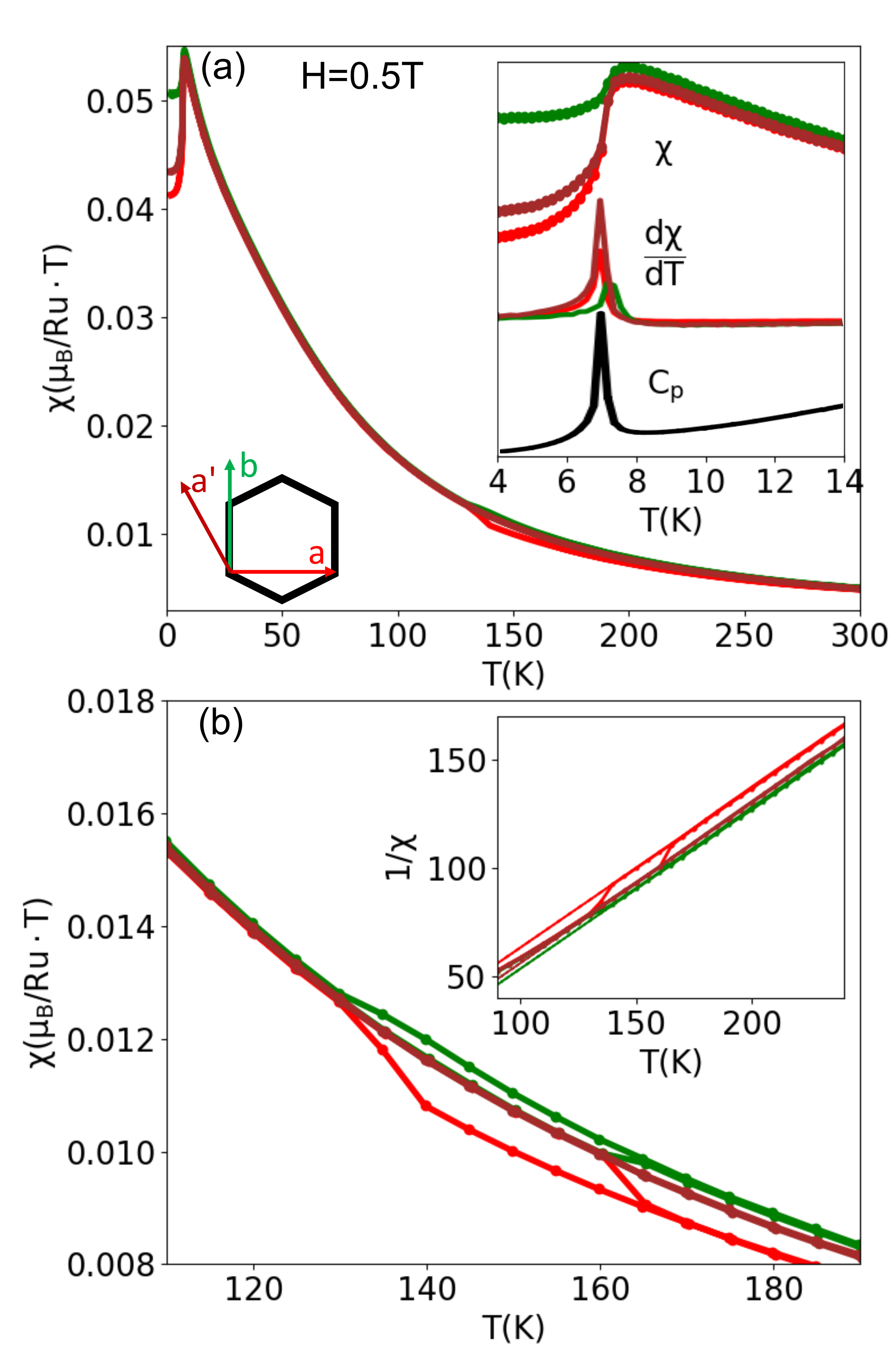}
    \caption{ (a) Temperature-dependent magnetic susceptibility with different in-plane field direction $\vec{a}$,$\vec{a'}$, and $\vec{b}$ where the directions are shown in Fig.~\ref{fig:structure}(a). Inset of (a) focuses on the magnetic transition. (b) Magnetic susceptibility close to the structural transition. The arrows denote heating and cooling directions. An identical range of hysteresis loops of 30K is observed for different high symmetry directions. However, the hysteresis behaviour is anisotropic. The susceptibility decreases in $\vec{a}$ while it increases in $\vec{b}$ and no change is seen in $\vec{a'}$. The inset shows the inverse susceptibility.}
    \label{fig:mag}
\end{figure}

\noindent
{\em In-plane Magnetic Anisotropy:} The local symmetry change is also corroborated by our magnetic susceptibility data. Figure~\ref{fig:mag} shows temperature dependent magnetic susceptibility $\chi(T)$ with a field applied in different high-symmetry directions in-plane: $\vec{a}$,$\vec{b}$ and $\vec{a}'$ as labelled in Fig.~\ref{fig:structure}. As shown in Fig.~\ref{fig:mag}(a), $\chi(T)$ is isotropic over a wide temperature range, except for two temperature regions. The inset of Fig.~\ref{fig:mag}(a) shows the region near the magnetic transition temperature $T_{N}$=7.2K, determined from the peaks in $d \chi/dT$ as well as in specific heat $C_p(T)$. The susceptibility along $a$, $\chi_a(T)$, is the smallest below $T_N$ as expected from the ordered moment along the $a$-axis (ignoring the tilt away from the honeycomb plane). The other region showing in-plane anisotropy is around the structural transition, which is shown in more detail in Fig.~\ref{fig:mag}(b).

Figure~\ref{fig:mag}(b) clearly shows that the magnetic susceptibility measured along the three directions is indistinguishable below 130~K but distinct above 160~K. The temperature hysteresis of the structural transition gives rise to somewhat complex behavior in between. The largest change in the susceptibility is observed for $\chi_a(T)$, while the change in the susceptibility is in the opposite sign for $\chi_{b}(T)$. $\chi_{a^\prime}(T)$ remains almost unchanged through the transition. To gain further understanding, we plot $\chi^{-1}(T)$ in the inset of Fig.~\ref{fig:mag}(b). We can see that the Curie constant, the slope of the inverse susceptibility, remains the same in all three directions at all temperatures (corresponding to a moment size of $2.46(6) \mu_B$). Therefore, the observed anisotropy can be attributed to different Weiss temperatures, shown as the vertical shifts in the plot. Since Curie constants are often associated with $g$-factors while Weiss temperatures depend strongly on the exchange interactions, we can conclude that the magnetic anisotropy in the high-temperature phase is caused by different magnetic interaction parameters.

Anisotropic magnetic susceptibility was investigated previously by Lampen-Kelley and coworkers, who observed that the in-plane magnetic susceptibility exhibits $C_2$ rotational symmetry, similar to the high-temperature data reported here \cite{lampenkelley2018}. The observed angle-dependence was explained using the high-temperature series expansion of the anisotropic $J-K-\Gamma$ model, allowing different interaction strengths between bonds along and perpendicular to the a-axis (i.e., zigzag direction). Our data could be quantitatively accounted for by using the same high-temperature expansion formula used in Ref.~\cite{lampenkelley2018} as shown in Supplemental Materials, although the Weiss temperatures are smaller than those obtained in Ref.~\cite{lampenkelley2018}. Of course, a major difference is the disappearance of the anisotropy in the low-temperature structure in our data, which presumably is due to the fact that our sample is mostly made up of a single twin domain of R$\bar{3}$. 
Our observation is consistent with the 6-fold symmetric angular dependence of specific heat reported in Yokoi et. al. \cite{yokoi2021}.

\noindent
{\em Discussion:} Let's first discuss the implication of the observed change in the in-plane magnetic anisotropy as a function of temperature. Both high- and low-temperature structures have inversion symmetry with respect to the center of the bond, which means that the $J-K-\Gamma$ model is still the minimal Hamiltonian for this material at low temperatures. We also note that additional small terms required to stabilize the zigzag ground state, such as the third-nearest neighbor interaction or the $\Gamma^\prime$ interaction due to trigonal crystal field, do not give rise to the magnetic anisotropy observed at high temperatures\cite{lampenkelley2018}. Therefore, observations of magnetic anisotropy can be only explained by the crystal structure explicitly breaking the local $C_3$ symmetry. Then, the magnetic anisotropy at low temperatures, reported in the literature, requires the presence of $C2/m$ structure.

This would be possible if the structural phase transition is incomplete and the high-temperature $C2/m$ structure coexists with the low temperature $R\bar{3}$ structure below the structural transition temperature. This seems to be the case in some low-quality samples as discussed in Ref.~\cite{Zhang2023}. Another possibility is that the $C2/m$ phase remains in the domain boundary region. Note that to go from $C2/m$ to $R\bar{3}$ structure, the upper layer should slide along the $-\frac{1}{3}\vec{a}+\frac{1}{3}\vec{b}+\vec{c}$ direction or the $-\frac{1}{3}\vec{a}-\frac{1}{3}\vec{b}+\vec{c}$ direction, resulting in the two twin domains discussed above. It is reasonable to assume that the twin domain boundary will remain in the $C2/m$ structure, and therefore heavily-twinned samples will show more of the residual $C2/m$ phase. This is consistent with the observation of isotropic magnetism in our sample with a minimal amount of twinning. A systematic investigation of the sample dependence will be reported elsewhere.

\noindent
{\em Conclusions:} We report our x-ray diffraction study of a structural phase transition in the high-quality (almost) twin-free $\alpha$-RuCl$_{3}$ crystals. $\alpha$-RuCl$_{3}$ goes through the structural transition from C2/m to R$\bar{3}$ structure, associated with the change in the stacking direction from the monoclinic a-axis to b-axis. We confirm that the bond-length anisotropy disappears in the R$\bar{3}$ phase, suggesting that the local symmetry of the honeycomb layer follows the global crystal symmetry. This is also supported by our observation of vanishing in-plane magnetic anisotropy in the low-temperature structure. Our study provides an unambiguous answer to the long-standing question about the low-temperature crystal structure of $\alpha$-RuCl$_{3}$, arguably the most promising candidate material for a Kitaev quantum spin liquid.   

\begin{acknowledgments}
We thank Jiefu Cen and Hae Young Kee for insightful discussions. Work at the University of Toronto was supported by the Natural Science and Engineering Research Council (NSERC) of Canada, Canadian Foundation for Innovation, and Ontario Research Fund. This work is based on research conducted at the Center for High-Energy X-ray Sciences (CHEXS), which is supported by the National Science Foundation (BIO, ENG and MPS Directorates) under award DMR-1829070. Part of the research described in this paper was performed at the Canadian Light Source, a national research facility of the University of Saskatchewan, which is supported by the Canada Foundation for Innovation (CFI), the Natural Sciences and Engineering Research Council (NSERC), the National Research Council (NRC), the Canadian Institutes of Health Research (CIHR), the Government of Saskatchewan, and the University of Saskatchewan. \end{acknowledgments}


%
%

%


\bibliography{RuCl3refs}

\begin{thebibliography}{46}%
\makeatletter
\providecommand \@ifxundefined [1]{%
 \@ifx{#1\undefined}
}%
\providecommand \@ifnum [1]{%
 \ifnum #1\expandafter \@firstoftwo
 \else \expandafter \@secondoftwo
 \fi
}%
\providecommand \@ifx [1]{%
 \ifx #1\expandafter \@firstoftwo
 \else \expandafter \@secondoftwo
 \fi
}%
\providecommand \natexlab [1]{#1}%
\providecommand \enquote  [1]{``#1''}%
\providecommand \bibnamefont  [1]{#1}%
\providecommand \bibfnamefont [1]{#1}%
\providecommand \citenamefont [1]{#1}%
\providecommand \href@noop [0]{\@secondoftwo}%
\providecommand \href [0]{\begingroup \@sanitize@url \@href}%
\providecommand \@href[1]{\@@startlink{#1}\@@href}%
\providecommand \@@href[1]{\endgroup#1\@@endlink}%
\providecommand \@sanitize@url [0]{\catcode `\\12\catcode `\$12\catcode
  `\&12\catcode `\#12\catcode `\^12\catcode `\_12\catcode `\%12\relax}%
\providecommand \@@startlink[1]{}%
\providecommand \@@endlink[0]{}%
\providecommand \url  [0]{\begingroup\@sanitize@url \@url }%
\providecommand \@url [1]{\endgroup\@href {#1}{\urlprefix }}%
\providecommand \urlprefix  [0]{URL }%
\providecommand \Eprint [0]{\href }%
\providecommand \doibase [0]{https://doi.org/}%
\providecommand \selectlanguage [0]{\@gobble}%
\providecommand \bibinfo  [0]{\@secondoftwo}%
\providecommand \bibfield  [0]{\@secondoftwo}%
\providecommand \translation [1]{[#1]}%
\providecommand \BibitemOpen [0]{}%
\providecommand \bibitemStop [0]{}%
\providecommand \bibitemNoStop [0]{.\EOS\space}%
\providecommand \EOS [0]{\spacefactor3000\relax}%
\providecommand \BibitemShut  [1]{\csname bibitem#1\endcsname}%
\let\auto@bib@innerbib\@empty
\bibitem [{\citenamefont {Plumb}\ \emph {et~al.}(2014)\citenamefont {Plumb},
  \citenamefont {Clancy}, \citenamefont {Sandilands}, \citenamefont {Shankar},
  \citenamefont {Hu}, \citenamefont {Burch}, \citenamefont {Kee},\ and\
  \citenamefont {Kim}}]{Plumb2014}%
  \BibitemOpen
  \bibfield  {author} {\bibinfo {author} {\bibfnamefont {K.~W.}\ \bibnamefont
  {Plumb}}, \bibinfo {author} {\bibfnamefont {J.~P.}\ \bibnamefont {Clancy}},
  \bibinfo {author} {\bibfnamefont {L.~J.}\ \bibnamefont {Sandilands}},
  \bibinfo {author} {\bibfnamefont {V.~V.}\ \bibnamefont {Shankar}}, \bibinfo
  {author} {\bibfnamefont {Y.~F.}\ \bibnamefont {Hu}}, \bibinfo {author}
  {\bibfnamefont {K.~S.}\ \bibnamefont {Burch}}, \bibinfo {author}
  {\bibfnamefont {H.-Y.}\ \bibnamefont {Kee}},\ and\ \bibinfo {author}
  {\bibfnamefont {Y.-J.}\ \bibnamefont {Kim}},\ }\bibfield  {title} {\bibinfo
  {title} {$\alpha$-{RuCl}$_3$: A spin-orbit assisted mott insulator on a
  honeycomb lattice},\ }\href {https://doi.org/10.1103/PhysRevB.90.041112}
  {\bibfield  {journal} {\bibinfo  {journal} {Phys. Rev. B}\ }\textbf {\bibinfo
  {volume} {90}},\ \bibinfo {pages} {041112} (\bibinfo {year}
  {2014})}\BibitemShut {NoStop}%
\bibitem [{\citenamefont {Sears}\ \emph {et~al.}(2015)\citenamefont {Sears},
  \citenamefont {Songvilay}, \citenamefont {Plumb}, \citenamefont {Clancy},
  \citenamefont {Qiu}, \citenamefont {Zhao}, \citenamefont {Parshall},\ and\
  \citenamefont {Kim}}]{sears15}%
  \BibitemOpen
  \bibfield  {author} {\bibinfo {author} {\bibfnamefont {J.~A.}\ \bibnamefont
  {Sears}}, \bibinfo {author} {\bibfnamefont {M.}~\bibnamefont {Songvilay}},
  \bibinfo {author} {\bibfnamefont {K.~W.}\ \bibnamefont {Plumb}}, \bibinfo
  {author} {\bibfnamefont {J.~P.}\ \bibnamefont {Clancy}}, \bibinfo {author}
  {\bibfnamefont {Y.}~\bibnamefont {Qiu}}, \bibinfo {author} {\bibfnamefont
  {Y.}~\bibnamefont {Zhao}}, \bibinfo {author} {\bibfnamefont {D.}~\bibnamefont
  {Parshall}},\ and\ \bibinfo {author} {\bibfnamefont {Y.-J.}\ \bibnamefont
  {Kim}},\ }\bibfield  {title} {\bibinfo {title} {Magnetic order in
  $\alpha$-{RuCl}$_3$: A honeycomb-lattice quantum magnet with strong
  spin-orbit coupling},\ }\href {https://doi.org/10.1103/PhysRevB.91.144420}
  {\bibfield  {journal} {\bibinfo  {journal} {Phys. Rev. B}\ }\textbf {\bibinfo
  {volume} {91}},\ \bibinfo {pages} {144420} (\bibinfo {year}
  {2015})}\BibitemShut {NoStop}%
\bibitem [{\citenamefont {Banerjee}\ \emph {et~al.}(2016)\citenamefont
  {Banerjee}, \citenamefont {Bridges}, \citenamefont {Yan}, \citenamefont
  {Aczel}, \citenamefont {Li}, \citenamefont {Stone}, \citenamefont {Granroth},
  \citenamefont {Lumsden}, \citenamefont {Yiu}, \citenamefont {Knolle},
  \citenamefont {Bhattacharjee}, \citenamefont {Kovrizhin}, \citenamefont
  {Moessner}, \citenamefont {Tennant}, \citenamefont {Mandrus},\ and\
  \citenamefont {Nagler}}]{banerjee16}%
  \BibitemOpen
  \bibfield  {author} {\bibinfo {author} {\bibfnamefont {A.}~\bibnamefont
  {Banerjee}}, \bibinfo {author} {\bibfnamefont {C.~A.}\ \bibnamefont
  {Bridges}}, \bibinfo {author} {\bibfnamefont {J.-Q.}\ \bibnamefont {Yan}},
  \bibinfo {author} {\bibfnamefont {A.~A.}\ \bibnamefont {Aczel}}, \bibinfo
  {author} {\bibfnamefont {L.}~\bibnamefont {Li}}, \bibinfo {author}
  {\bibfnamefont {M.~B.}\ \bibnamefont {Stone}}, \bibinfo {author}
  {\bibfnamefont {G.~E.}\ \bibnamefont {Granroth}}, \bibinfo {author}
  {\bibfnamefont {M.~D.}\ \bibnamefont {Lumsden}}, \bibinfo {author}
  {\bibfnamefont {Y.}~\bibnamefont {Yiu}}, \bibinfo {author} {\bibfnamefont
  {J.}~\bibnamefont {Knolle}}, \bibinfo {author} {\bibfnamefont
  {S.}~\bibnamefont {Bhattacharjee}}, \bibinfo {author} {\bibfnamefont {D.~L.}\
  \bibnamefont {Kovrizhin}}, \bibinfo {author} {\bibfnamefont {R.}~\bibnamefont
  {Moessner}}, \bibinfo {author} {\bibfnamefont {D.~A.}\ \bibnamefont
  {Tennant}}, \bibinfo {author} {\bibfnamefont {D.~G.}\ \bibnamefont
  {Mandrus}},\ and\ \bibinfo {author} {\bibfnamefont {S.~E.}\ \bibnamefont
  {Nagler}},\ }\bibfield  {title} {\bibinfo {title} {"proximate kitaev quantum
  spin liquid behaviour in a honeycomb magnet"},\ }\href
  {http://dx.doi.org/10.1038/nmat4604} {\bibfield  {journal} {\bibinfo
  {journal} {Nat. Mater.}\ }\textbf {\bibinfo {volume} {15}},\ \bibinfo {pages}
  {733} (\bibinfo {year} {2016})},\ \bibinfo {note} {article}\BibitemShut
  {NoStop}%
\bibitem [{\citenamefont {Do}\ \emph {et~al.}(2017)\citenamefont {Do},
  \citenamefont {Park}, \citenamefont {Yoshitake}, \citenamefont {Nasu},
  \citenamefont {Motome}, \citenamefont {Kwon}, \citenamefont {Adroja},
  \citenamefont {Voneshen}, \citenamefont {Kim}, \citenamefont {Jang},
  \citenamefont {Park}, \citenamefont {Choi},\ and\ \citenamefont
  {Ji}}]{Do2017}%
  \BibitemOpen
  \bibfield  {author} {\bibinfo {author} {\bibfnamefont {S.-H.}\ \bibnamefont
  {Do}}, \bibinfo {author} {\bibfnamefont {S.-Y.}\ \bibnamefont {Park}},
  \bibinfo {author} {\bibfnamefont {J.}~\bibnamefont {Yoshitake}}, \bibinfo
  {author} {\bibfnamefont {J.}~\bibnamefont {Nasu}}, \bibinfo {author}
  {\bibfnamefont {Y.}~\bibnamefont {Motome}}, \bibinfo {author} {\bibfnamefont
  {Y.~S.}\ \bibnamefont {Kwon}}, \bibinfo {author} {\bibfnamefont {D.~T.}\
  \bibnamefont {Adroja}}, \bibinfo {author} {\bibfnamefont {D.~J.}\
  \bibnamefont {Voneshen}}, \bibinfo {author} {\bibfnamefont {K.}~\bibnamefont
  {Kim}}, \bibinfo {author} {\bibfnamefont {T.-H.}\ \bibnamefont {Jang}},
  \bibinfo {author} {\bibfnamefont {J.-H.}\ \bibnamefont {Park}}, \bibinfo
  {author} {\bibfnamefont {K.-Y.}\ \bibnamefont {Choi}},\ and\ \bibinfo
  {author} {\bibfnamefont {S.}~\bibnamefont {Ji}},\ }\bibfield  {title}
  {\bibinfo {title} {Majorana fermions in the kitaev quantum spin system
  $\alpha$-{RuCl}$_3$},\ }\href {https://doi.org/10.1038/nphys4264} {\bibfield
  {journal} {\bibinfo  {journal} {Nature Physics}\ }\textbf {\bibinfo {volume}
  {13}},\ \bibinfo {pages} {1079} (\bibinfo {year} {2017})}\BibitemShut
  {NoStop}%
\bibitem [{\citenamefont {Banerjee}\ \emph {et~al.}(2018)\citenamefont
  {Banerjee}, \citenamefont {Lampen-Kelley}, \citenamefont {Knolle},
  \citenamefont {Balz}, \citenamefont {Aczel}, \citenamefont {Winn},
  \citenamefont {Liu}, \citenamefont {Pajerowski}, \citenamefont {Yan},
  \citenamefont {Bridges}, \citenamefont {Savici}, \citenamefont {Chakoumakos},
  \citenamefont {Lumsden}, \citenamefont {Tennant}, \citenamefont {Moessner},
  \citenamefont {Mandrus},\ and\ \citenamefont {Nagler}}]{Banerjee2018}%
  \BibitemOpen
  \bibfield  {author} {\bibinfo {author} {\bibfnamefont {A.}~\bibnamefont
  {Banerjee}}, \bibinfo {author} {\bibfnamefont {P.}~\bibnamefont
  {Lampen-Kelley}}, \bibinfo {author} {\bibfnamefont {J.}~\bibnamefont
  {Knolle}}, \bibinfo {author} {\bibfnamefont {C.}~\bibnamefont {Balz}},
  \bibinfo {author} {\bibfnamefont {A.~A.}\ \bibnamefont {Aczel}}, \bibinfo
  {author} {\bibfnamefont {B.}~\bibnamefont {Winn}}, \bibinfo {author}
  {\bibfnamefont {Y.}~\bibnamefont {Liu}}, \bibinfo {author} {\bibfnamefont
  {D.}~\bibnamefont {Pajerowski}}, \bibinfo {author} {\bibfnamefont
  {J.}~\bibnamefont {Yan}}, \bibinfo {author} {\bibfnamefont {C.~A.}\
  \bibnamefont {Bridges}}, \bibinfo {author} {\bibfnamefont {A.~T.}\
  \bibnamefont {Savici}}, \bibinfo {author} {\bibfnamefont {B.~C.}\
  \bibnamefont {Chakoumakos}}, \bibinfo {author} {\bibfnamefont {M.~D.}\
  \bibnamefont {Lumsden}}, \bibinfo {author} {\bibfnamefont {D.~A.}\
  \bibnamefont {Tennant}}, \bibinfo {author} {\bibfnamefont {R.}~\bibnamefont
  {Moessner}}, \bibinfo {author} {\bibfnamefont {D.~G.}\ \bibnamefont
  {Mandrus}},\ and\ \bibinfo {author} {\bibfnamefont {S.~E.}\ \bibnamefont
  {Nagler}},\ }\bibfield  {title} {\bibinfo {title} {Excitations in the
  field-induced quantum spin liquid state of $\alpha$-rucl3},\ }\href
  {https://doi.org/10.1038/s41535-018-0079-2} {\bibfield  {journal} {\bibinfo
  {journal} {npj Quantum Materials}\ }\textbf {\bibinfo {volume} {3}},\
  \bibinfo {pages} {8} (\bibinfo {year} {2018})}\BibitemShut {NoStop}%
\bibitem [{\citenamefont {Sandilands}\ \emph {et~al.}(2015)\citenamefont
  {Sandilands}, \citenamefont {Tian}, \citenamefont {Plumb}, \citenamefont
  {Kim},\ and\ \citenamefont {Burch}}]{sandilands15}%
  \BibitemOpen
  \bibfield  {author} {\bibinfo {author} {\bibfnamefont {L.~J.}\ \bibnamefont
  {Sandilands}}, \bibinfo {author} {\bibfnamefont {Y.}~\bibnamefont {Tian}},
  \bibinfo {author} {\bibfnamefont {K.~W.}\ \bibnamefont {Plumb}}, \bibinfo
  {author} {\bibfnamefont {Y.-J.}\ \bibnamefont {Kim}},\ and\ \bibinfo {author}
  {\bibfnamefont {K.~S.}\ \bibnamefont {Burch}},\ }\bibfield  {title} {\bibinfo
  {title} {Scattering continuum and possible fractionalized excitations in
  $\alpha$-{RuCl}$_3$},\ }\href
  {https://doi.org/10.1103/PhysRevLett.114.147201} {\bibfield  {journal}
  {\bibinfo  {journal} {Phys. Rev. Lett.}\ }\textbf {\bibinfo {volume} {114}},\
  \bibinfo {pages} {147201} (\bibinfo {year} {2015})}\BibitemShut {NoStop}%
\bibitem [{\citenamefont {Majumder}\ \emph {et~al.}(2015)\citenamefont
  {Majumder}, \citenamefont {Schmidt}, \citenamefont {Rosner}, \citenamefont
  {Tsirlin}, \citenamefont {Yasuoka},\ and\ \citenamefont
  {Baenitz}}]{majumder15}%
  \BibitemOpen
  \bibfield  {author} {\bibinfo {author} {\bibfnamefont {M.}~\bibnamefont
  {Majumder}}, \bibinfo {author} {\bibfnamefont {M.}~\bibnamefont {Schmidt}},
  \bibinfo {author} {\bibfnamefont {H.}~\bibnamefont {Rosner}}, \bibinfo
  {author} {\bibfnamefont {A.~A.}\ \bibnamefont {Tsirlin}}, \bibinfo {author}
  {\bibfnamefont {H.}~\bibnamefont {Yasuoka}},\ and\ \bibinfo {author}
  {\bibfnamefont {M.}~\bibnamefont {Baenitz}},\ }\bibfield  {title} {\bibinfo
  {title} {Anisotropic ${\mathrm{ru}}^{3+} 4{d}^{5}$ magnetism in the
  $\alpha$-{RuCl}$_3$ honeycomb system: Susceptibility, specific heat, and
  zero-field nmr},\ }\href {https://doi.org/10.1103/PhysRevB.91.180401}
  {\bibfield  {journal} {\bibinfo  {journal} {Phys. Rev. B}\ }\textbf {\bibinfo
  {volume} {91}},\ \bibinfo {pages} {180401} (\bibinfo {year}
  {2015})}\BibitemShut {NoStop}%
\bibitem [{\citenamefont {Kubota}\ \emph {et~al.}(2015)\citenamefont {Kubota},
  \citenamefont {Tanaka}, \citenamefont {Ono}, \citenamefont {Narumi},\ and\
  \citenamefont {Kindo}}]{kubota15}%
  \BibitemOpen
  \bibfield  {author} {\bibinfo {author} {\bibfnamefont {Y.}~\bibnamefont
  {Kubota}}, \bibinfo {author} {\bibfnamefont {H.}~\bibnamefont {Tanaka}},
  \bibinfo {author} {\bibfnamefont {T.}~\bibnamefont {Ono}}, \bibinfo {author}
  {\bibfnamefont {Y.}~\bibnamefont {Narumi}},\ and\ \bibinfo {author}
  {\bibfnamefont {K.}~\bibnamefont {Kindo}},\ }\bibfield  {title} {\bibinfo
  {title} {Successive magnetic phase transitions in $\alpha$-{RuCl}$_3$:
  Xy-like frustrated magnet on the honeycomb lattice},\ }\href
  {https://doi.org/10.1103/PhysRevB.91.094422} {\bibfield  {journal} {\bibinfo
  {journal} {Phys. Rev. B}\ }\textbf {\bibinfo {volume} {91}},\ \bibinfo
  {pages} {094422} (\bibinfo {year} {2015})}\BibitemShut {NoStop}%
\bibitem [{\citenamefont {Johnson}\ \emph {et~al.}(2015)\citenamefont
  {Johnson}, \citenamefont {Williams}, \citenamefont {Haghighirad},
  \citenamefont {Singleton}, \citenamefont {Zapf}, \citenamefont {Manuel},
  \citenamefont {Mazin}, \citenamefont {Li}, \citenamefont {Jeschke},
  \citenamefont {Valent\'{\i}},\ and\ \citenamefont {Coldea}}]{johnson15}%
  \BibitemOpen
  \bibfield  {author} {\bibinfo {author} {\bibfnamefont {R.~D.}\ \bibnamefont
  {Johnson}}, \bibinfo {author} {\bibfnamefont {S.~C.}\ \bibnamefont
  {Williams}}, \bibinfo {author} {\bibfnamefont {A.~A.}\ \bibnamefont
  {Haghighirad}}, \bibinfo {author} {\bibfnamefont {J.}~\bibnamefont
  {Singleton}}, \bibinfo {author} {\bibfnamefont {V.}~\bibnamefont {Zapf}},
  \bibinfo {author} {\bibfnamefont {P.}~\bibnamefont {Manuel}}, \bibinfo
  {author} {\bibfnamefont {I.~I.}\ \bibnamefont {Mazin}}, \bibinfo {author}
  {\bibfnamefont {Y.}~\bibnamefont {Li}}, \bibinfo {author} {\bibfnamefont
  {H.~O.}\ \bibnamefont {Jeschke}}, \bibinfo {author} {\bibfnamefont
  {R.}~\bibnamefont {Valent\'{\i}}},\ and\ \bibinfo {author} {\bibfnamefont
  {R.}~\bibnamefont {Coldea}},\ }\bibfield  {title} {\bibinfo {title}
  {Monoclinic crystal structure of $\alpha$-{RuCl}$_3$ and the zigzag
  antiferromagnetic ground state},\ }\href
  {https://doi.org/10.1103/PhysRevB.92.235119} {\bibfield  {journal} {\bibinfo
  {journal} {Phys. Rev. B}\ }\textbf {\bibinfo {volume} {92}},\ \bibinfo
  {pages} {235119} (\bibinfo {year} {2015})}\BibitemShut {NoStop}%
\bibitem [{\citenamefont {Cao}\ \emph {et~al.}(2016)\citenamefont {Cao},
  \citenamefont {Banerjee}, \citenamefont {Yan}, \citenamefont {Bridges},
  \citenamefont {Lumsden}, \citenamefont {Mandrus}, \citenamefont {Tennant},
  \citenamefont {Chakoumakos},\ and\ \citenamefont {Nagler}}]{cao16}%
  \BibitemOpen
  \bibfield  {author} {\bibinfo {author} {\bibfnamefont {H.~B.}\ \bibnamefont
  {Cao}}, \bibinfo {author} {\bibfnamefont {A.}~\bibnamefont {Banerjee}},
  \bibinfo {author} {\bibfnamefont {J.-Q.}\ \bibnamefont {Yan}}, \bibinfo
  {author} {\bibfnamefont {C.~A.}\ \bibnamefont {Bridges}}, \bibinfo {author}
  {\bibfnamefont {M.~D.}\ \bibnamefont {Lumsden}}, \bibinfo {author}
  {\bibfnamefont {D.~G.}\ \bibnamefont {Mandrus}}, \bibinfo {author}
  {\bibfnamefont {D.~A.}\ \bibnamefont {Tennant}}, \bibinfo {author}
  {\bibfnamefont {B.~C.}\ \bibnamefont {Chakoumakos}},\ and\ \bibinfo {author}
  {\bibfnamefont {S.~E.}\ \bibnamefont {Nagler}},\ }\bibfield  {title}
  {\bibinfo {title} {Low-temperature crystal and magnetic structure of
  $\alpha$-{RuCl}$_3$},\ }\href {https://doi.org/10.1103/PhysRevB.93.134423}
  {\bibfield  {journal} {\bibinfo  {journal} {Phys. Rev. B}\ }\textbf {\bibinfo
  {volume} {93}},\ \bibinfo {pages} {134423} (\bibinfo {year}
  {2016})}\BibitemShut {NoStop}%
\bibitem [{\citenamefont {Leahy}\ \emph {et~al.}(2017)\citenamefont {Leahy},
  \citenamefont {Pocs}, \citenamefont {Siegfried}, \citenamefont {Graf},
  \citenamefont {Do}, \citenamefont {Choi}, \citenamefont {Normand},\ and\
  \citenamefont {Lee}}]{leahy2017}%
  \BibitemOpen
  \bibfield  {author} {\bibinfo {author} {\bibfnamefont {I.~A.}\ \bibnamefont
  {Leahy}}, \bibinfo {author} {\bibfnamefont {C.~A.}\ \bibnamefont {Pocs}},
  \bibinfo {author} {\bibfnamefont {P.~E.}\ \bibnamefont {Siegfried}}, \bibinfo
  {author} {\bibfnamefont {D.}~\bibnamefont {Graf}}, \bibinfo {author}
  {\bibfnamefont {S.-H.}\ \bibnamefont {Do}}, \bibinfo {author} {\bibfnamefont
  {K.-Y.}\ \bibnamefont {Choi}}, \bibinfo {author} {\bibfnamefont
  {B.}~\bibnamefont {Normand}},\ and\ \bibinfo {author} {\bibfnamefont
  {M.}~\bibnamefont {Lee}},\ }\bibfield  {title} {\bibinfo {title} {Anomalous
  thermal conductivity and magnetic torque response in the honeycomb magnet
  $\alpha$-{RuCl}$_3$},\ }\href
  {https://doi.org/10.1103/PhysRevLett.118.187203} {\bibfield  {journal}
  {\bibinfo  {journal} {Phys. Rev. Lett.}\ }\textbf {\bibinfo {volume} {118}},\
  \bibinfo {pages} {187203} (\bibinfo {year} {2017})}\BibitemShut {NoStop}%
\bibitem [{\citenamefont {Baek}\ \emph {et~al.}(2017)\citenamefont {Baek},
  \citenamefont {Do}, \citenamefont {Choi}, \citenamefont {Kwon}, \citenamefont
  {Wolter}, \citenamefont {Nishimoto}, \citenamefont {van~den Brink},\ and\
  \citenamefont {B\"uchner}}]{baek2017}%
  \BibitemOpen
  \bibfield  {author} {\bibinfo {author} {\bibfnamefont {S.-H.}\ \bibnamefont
  {Baek}}, \bibinfo {author} {\bibfnamefont {S.-H.}\ \bibnamefont {Do}},
  \bibinfo {author} {\bibfnamefont {K.-Y.}\ \bibnamefont {Choi}}, \bibinfo
  {author} {\bibfnamefont {Y.~S.}\ \bibnamefont {Kwon}}, \bibinfo {author}
  {\bibfnamefont {A.~U.~B.}\ \bibnamefont {Wolter}}, \bibinfo {author}
  {\bibfnamefont {S.}~\bibnamefont {Nishimoto}}, \bibinfo {author}
  {\bibfnamefont {J.}~\bibnamefont {van~den Brink}},\ and\ \bibinfo {author}
  {\bibfnamefont {B.}~\bibnamefont {B\"uchner}},\ }\bibfield  {title} {\bibinfo
  {title} {Evidence for a field-induced quantum spin liquid in
  ${\alpha}$-{RuCl}$_3$},\ }\href
  {https://doi.org/10.1103/PhysRevLett.119.037201} {\bibfield  {journal}
  {\bibinfo  {journal} {Phys. Rev. Lett.}\ }\textbf {\bibinfo {volume} {119}},\
  \bibinfo {pages} {037201} (\bibinfo {year} {2017})}\BibitemShut {NoStop}%
\bibitem [{\citenamefont {Sears}\ \emph {et~al.}(2017)\citenamefont {Sears},
  \citenamefont {Zhao}, \citenamefont {Xu}, \citenamefont {Lynn},\ and\
  \citenamefont {Kim}}]{Sears2017}%
  \BibitemOpen
  \bibfield  {author} {\bibinfo {author} {\bibfnamefont {J.~A.}\ \bibnamefont
  {Sears}}, \bibinfo {author} {\bibfnamefont {Y.}~\bibnamefont {Zhao}},
  \bibinfo {author} {\bibfnamefont {Z.}~\bibnamefont {Xu}}, \bibinfo {author}
  {\bibfnamefont {J.~W.}\ \bibnamefont {Lynn}},\ and\ \bibinfo {author}
  {\bibfnamefont {Y.-J.}\ \bibnamefont {Kim}},\ }\bibfield  {title} {\bibinfo
  {title} {Phase diagram of ${\alpha}$-{RuCl}$_{3}$ in an in-plane magnetic
  field},\ }\href {https://doi.org/10.1103/PhysRevB.95.180411} {\bibfield
  {journal} {\bibinfo  {journal} {Phys. Rev. B}\ }\textbf {\bibinfo {volume}
  {95}},\ \bibinfo {pages} {180411} (\bibinfo {year} {2017})}\BibitemShut
  {NoStop}%
\bibitem [{\citenamefont {Wolter}\ \emph {et~al.}(2017)\citenamefont {Wolter},
  \citenamefont {Corredor}, \citenamefont {Janssen}, \citenamefont {Nenkov},
  \citenamefont {Sch\"onecker}, \citenamefont {Do}, \citenamefont {Choi},
  \citenamefont {Albrecht}, \citenamefont {Hunger}, \citenamefont {Doert},
  \citenamefont {Vojta},\ and\ \citenamefont {B\"uchner}}]{wolter2017}%
  \BibitemOpen
  \bibfield  {author} {\bibinfo {author} {\bibfnamefont {A.~U.~B.}\
  \bibnamefont {Wolter}}, \bibinfo {author} {\bibfnamefont {L.~T.}\
  \bibnamefont {Corredor}}, \bibinfo {author} {\bibfnamefont {L.}~\bibnamefont
  {Janssen}}, \bibinfo {author} {\bibfnamefont {K.}~\bibnamefont {Nenkov}},
  \bibinfo {author} {\bibfnamefont {S.}~\bibnamefont {Sch\"onecker}}, \bibinfo
  {author} {\bibfnamefont {S.-H.}\ \bibnamefont {Do}}, \bibinfo {author}
  {\bibfnamefont {K.-Y.}\ \bibnamefont {Choi}}, \bibinfo {author}
  {\bibfnamefont {R.}~\bibnamefont {Albrecht}}, \bibinfo {author}
  {\bibfnamefont {J.}~\bibnamefont {Hunger}}, \bibinfo {author} {\bibfnamefont
  {T.}~\bibnamefont {Doert}}, \bibinfo {author} {\bibfnamefont
  {M.}~\bibnamefont {Vojta}},\ and\ \bibinfo {author} {\bibfnamefont
  {B.}~\bibnamefont {B\"uchner}},\ }\bibfield  {title} {\bibinfo {title}
  {Field-induced quantum criticality in the kitaev system
  ${\alpha}$-{RuCl}$_3$},\ }\href {https://doi.org/10.1103/PhysRevB.96.041405}
  {\bibfield  {journal} {\bibinfo  {journal} {Phys. Rev. B}\ }\textbf {\bibinfo
  {volume} {96}},\ \bibinfo {pages} {041405} (\bibinfo {year}
  {2017})}\BibitemShut {NoStop}%
\bibitem [{\citenamefont {Zheng}\ \emph {et~al.}(2017)\citenamefont {Zheng},
  \citenamefont {Ran}, \citenamefont {Li}, \citenamefont {Wang}, \citenamefont
  {Wang}, \citenamefont {Liu}, \citenamefont {Liu}, \citenamefont {Normand},
  \citenamefont {Wen},\ and\ \citenamefont {Yu}}]{zheng2017}%
  \BibitemOpen
  \bibfield  {author} {\bibinfo {author} {\bibfnamefont {J.}~\bibnamefont
  {Zheng}}, \bibinfo {author} {\bibfnamefont {K.}~\bibnamefont {Ran}}, \bibinfo
  {author} {\bibfnamefont {T.}~\bibnamefont {Li}}, \bibinfo {author}
  {\bibfnamefont {J.}~\bibnamefont {Wang}}, \bibinfo {author} {\bibfnamefont
  {P.}~\bibnamefont {Wang}}, \bibinfo {author} {\bibfnamefont {B.}~\bibnamefont
  {Liu}}, \bibinfo {author} {\bibfnamefont {Z.-X.}\ \bibnamefont {Liu}},
  \bibinfo {author} {\bibfnamefont {B.}~\bibnamefont {Normand}}, \bibinfo
  {author} {\bibfnamefont {J.}~\bibnamefont {Wen}},\ and\ \bibinfo {author}
  {\bibfnamefont {W.}~\bibnamefont {Yu}},\ }\bibfield  {title} {\bibinfo
  {title} {Gapless spin excitations in the field-induced quantum spin liquid
  phase of ${\alpha}$-{RuCl}$_3$},\ }\href
  {https://doi.org/10.1103/PhysRevLett.119.227208} {\bibfield  {journal}
  {\bibinfo  {journal} {Phys. Rev. Lett.}\ }\textbf {\bibinfo {volume} {119}},\
  \bibinfo {pages} {227208} (\bibinfo {year} {2017})}\BibitemShut {NoStop}%
\bibitem [{\citenamefont {Kasahara}\ \emph {et~al.}(2018)\citenamefont
  {Kasahara}, \citenamefont {Ohnishi}, \citenamefont {Mizukami}, \citenamefont
  {Tanaka}, \citenamefont {Ma}, \citenamefont {Sugii}, \citenamefont {Kurita},
  \citenamefont {Tanaka}, \citenamefont {Nasu}, \citenamefont {Motome},
  \citenamefont {Shibauchi},\ and\ \citenamefont {Matsuda}}]{Kasahara2018}%
  \BibitemOpen
  \bibfield  {author} {\bibinfo {author} {\bibfnamefont {Y.}~\bibnamefont
  {Kasahara}}, \bibinfo {author} {\bibfnamefont {T.}~\bibnamefont {Ohnishi}},
  \bibinfo {author} {\bibfnamefont {Y.}~\bibnamefont {Mizukami}}, \bibinfo
  {author} {\bibfnamefont {O.}~\bibnamefont {Tanaka}}, \bibinfo {author}
  {\bibfnamefont {S.}~\bibnamefont {Ma}}, \bibinfo {author} {\bibfnamefont
  {K.}~\bibnamefont {Sugii}}, \bibinfo {author} {\bibfnamefont
  {N.}~\bibnamefont {Kurita}}, \bibinfo {author} {\bibfnamefont
  {H.}~\bibnamefont {Tanaka}}, \bibinfo {author} {\bibfnamefont
  {J.}~\bibnamefont {Nasu}}, \bibinfo {author} {\bibfnamefont {Y.}~\bibnamefont
  {Motome}}, \bibinfo {author} {\bibfnamefont {T.}~\bibnamefont {Shibauchi}},\
  and\ \bibinfo {author} {\bibfnamefont {Y.}~\bibnamefont {Matsuda}},\
  }\bibfield  {title} {\bibinfo {title} {"majorana quantization and
  half-integer thermal quantum hall effect in a kitaev spin liquid"},\ }\href
  {https://doi.org/10.1038/s41586-018-0274-0} {\bibfield  {journal} {\bibinfo
  {journal} {Nature}\ }\textbf {\bibinfo {volume} {559}},\ \bibinfo {pages}
  {227} (\bibinfo {year} {2018})}\BibitemShut {NoStop}%
\bibitem [{\citenamefont {Lampen-Kelley}\ \emph {et~al.}(2018)\citenamefont
  {Lampen-Kelley}, \citenamefont {Rachel}, \citenamefont {Reuther},
  \citenamefont {Yan}, \citenamefont {Banerjee}, \citenamefont {Bridges},
  \citenamefont {Cao}, \citenamefont {Nagler},\ and\ \citenamefont
  {Mandrus}}]{lampenkelley2018}%
  \BibitemOpen
  \bibfield  {author} {\bibinfo {author} {\bibfnamefont {P.}~\bibnamefont
  {Lampen-Kelley}}, \bibinfo {author} {\bibfnamefont {S.}~\bibnamefont
  {Rachel}}, \bibinfo {author} {\bibfnamefont {J.}~\bibnamefont {Reuther}},
  \bibinfo {author} {\bibfnamefont {J.-Q.}\ \bibnamefont {Yan}}, \bibinfo
  {author} {\bibfnamefont {A.}~\bibnamefont {Banerjee}}, \bibinfo {author}
  {\bibfnamefont {C.~A.}\ \bibnamefont {Bridges}}, \bibinfo {author}
  {\bibfnamefont {H.~B.}\ \bibnamefont {Cao}}, \bibinfo {author} {\bibfnamefont
  {S.~E.}\ \bibnamefont {Nagler}},\ and\ \bibinfo {author} {\bibfnamefont
  {D.}~\bibnamefont {Mandrus}},\ }\bibfield  {title} {\bibinfo {title}
  {Anisotropic susceptibilities in the honeycomb {Kitaev} system
  $\alpha$-{RuCl}$_3$},\ }\href {https://doi.org/10.1103/PhysRevB.98.100403}
  {\bibfield  {journal} {\bibinfo  {journal} {Phys. Rev. B}\ }\textbf {\bibinfo
  {volume} {98}},\ \bibinfo {pages} {100403} (\bibinfo {year}
  {2018})}\BibitemShut {NoStop}%
\bibitem [{\citenamefont {Modic}\ \emph
  {et~al.}(2018{\natexlab{a}})\citenamefont {Modic}, \citenamefont {Bachmann},
  \citenamefont {Ramshaw}, \citenamefont {Arnold}, \citenamefont {Shirer},
  \citenamefont {Estry}, \citenamefont {Betts}, \citenamefont {Ghimire},
  \citenamefont {Bauer}, \citenamefont {Schmidt}, \citenamefont {Baenitz},
  \citenamefont {Svanidze}, \citenamefont {McDonald}, \citenamefont
  {Shekhter},\ and\ \citenamefont {Moll}}]{modic2018NC}%
  \BibitemOpen
  \bibfield  {author} {\bibinfo {author} {\bibfnamefont {K.~A.}\ \bibnamefont
  {Modic}}, \bibinfo {author} {\bibfnamefont {M.~D.}\ \bibnamefont {Bachmann}},
  \bibinfo {author} {\bibfnamefont {B.~J.}\ \bibnamefont {Ramshaw}}, \bibinfo
  {author} {\bibfnamefont {F.}~\bibnamefont {Arnold}}, \bibinfo {author}
  {\bibfnamefont {K.~R.}\ \bibnamefont {Shirer}}, \bibinfo {author}
  {\bibfnamefont {A.}~\bibnamefont {Estry}}, \bibinfo {author} {\bibfnamefont
  {J.~B.}\ \bibnamefont {Betts}}, \bibinfo {author} {\bibfnamefont {N.~J.}\
  \bibnamefont {Ghimire}}, \bibinfo {author} {\bibfnamefont {E.~D.}\
  \bibnamefont {Bauer}}, \bibinfo {author} {\bibfnamefont {M.}~\bibnamefont
  {Schmidt}}, \bibinfo {author} {\bibfnamefont {M.}~\bibnamefont {Baenitz}},
  \bibinfo {author} {\bibfnamefont {E.}~\bibnamefont {Svanidze}}, \bibinfo
  {author} {\bibfnamefont {R.~D.}\ \bibnamefont {McDonald}}, \bibinfo {author}
  {\bibfnamefont {A.}~\bibnamefont {Shekhter}},\ and\ \bibinfo {author}
  {\bibfnamefont {P.~J.~W.}\ \bibnamefont {Moll}},\ }\bibfield  {title}
  {\bibinfo {title} {Resonant torsion magnetometry in anisotropic quantum
  materials},\ }\href {https://doi.org/10.1038/s41467-018-06412-w} {\bibfield
  {journal} {\bibinfo  {journal} {Nat. Commun.}\ }\textbf {\bibinfo {volume}
  {9}},\ \bibinfo {pages} {3975} (\bibinfo {year}
  {2018}{\natexlab{a}})}\BibitemShut {NoStop}%
\bibitem [{\citenamefont {Yokoi}\ \emph {et~al.}(2021)\citenamefont {Yokoi},
  \citenamefont {Ma}, \citenamefont {Kasahara}, \citenamefont {Kasahara},
  \citenamefont {Shibauchi}, \citenamefont {Kurita}, \citenamefont {Tanaka},
  \citenamefont {Nasu}, \citenamefont {Motome}, \citenamefont {Hickey},
  \citenamefont {Trebst},\ and\ \citenamefont {Matsuda}}]{yokoi2021}%
  \BibitemOpen
  \bibfield  {author} {\bibinfo {author} {\bibfnamefont {T.}~\bibnamefont
  {Yokoi}}, \bibinfo {author} {\bibfnamefont {S.}~\bibnamefont {Ma}}, \bibinfo
  {author} {\bibfnamefont {Y.}~\bibnamefont {Kasahara}}, \bibinfo {author}
  {\bibfnamefont {S.}~\bibnamefont {Kasahara}}, \bibinfo {author}
  {\bibfnamefont {T.}~\bibnamefont {Shibauchi}}, \bibinfo {author}
  {\bibfnamefont {N.}~\bibnamefont {Kurita}}, \bibinfo {author} {\bibfnamefont
  {H.}~\bibnamefont {Tanaka}}, \bibinfo {author} {\bibfnamefont
  {J.}~\bibnamefont {Nasu}}, \bibinfo {author} {\bibfnamefont {Y.}~\bibnamefont
  {Motome}}, \bibinfo {author} {\bibfnamefont {C.}~\bibnamefont {Hickey}},
  \bibinfo {author} {\bibfnamefont {S.}~\bibnamefont {Trebst}},\ and\ \bibinfo
  {author} {\bibfnamefont {Y.}~\bibnamefont {Matsuda}},\ }\bibfield  {title}
  {\bibinfo {title} {Half-integer quantized anomalous thermal {Hall} effect in
  the {Kitaev} material candidate $\alpha$-{RuCl}$_3$},\ }\href
  {https://doi.org/10.1126/science.aay5551} {\bibfield  {journal} {\bibinfo
  {journal} {Science}\ }\textbf {\bibinfo {volume} {373}},\ \bibinfo {pages}
  {568} (\bibinfo {year} {2021})}\BibitemShut {NoStop}%
\bibitem [{\citenamefont {Bruin}\ \emph {et~al.}(2022)\citenamefont {Bruin},
  \citenamefont {Claus}, \citenamefont {Matsumoto}, \citenamefont {Kurita},
  \citenamefont {Tanaka},\ and\ \citenamefont {Takagi}}]{bruin2022}%
  \BibitemOpen
  \bibfield  {author} {\bibinfo {author} {\bibfnamefont {J.~A.~N.}\
  \bibnamefont {Bruin}}, \bibinfo {author} {\bibfnamefont {R.~R.}\ \bibnamefont
  {Claus}}, \bibinfo {author} {\bibfnamefont {Y.}~\bibnamefont {Matsumoto}},
  \bibinfo {author} {\bibfnamefont {N.}~\bibnamefont {Kurita}}, \bibinfo
  {author} {\bibfnamefont {H.}~\bibnamefont {Tanaka}},\ and\ \bibinfo {author}
  {\bibfnamefont {H.}~\bibnamefont {Takagi}},\ }\bibfield  {title} {\bibinfo
  {title} {Robustness of the thermal {Hall} effect close to half-quantization
  in $\alpha$-{RuCl}$_3$},\ }\href {https://doi.org/10.1038/s41567-021-01501-y}
  {\bibfield  {journal} {\bibinfo  {journal} {Nat. Phys.}\ }\textbf {\bibinfo
  {volume} {18}},\ \bibinfo {pages} {401} (\bibinfo {year} {2022})}\BibitemShut
  {NoStop}%
\bibitem [{\citenamefont {Lefran\ifmmode~\mbox{\c{c}}\else \c{c}\fi{}ois}\
  \emph {et~al.}(2022)\citenamefont {Lefran\ifmmode~\mbox{\c{c}}\else
  \c{c}\fi{}ois}, \citenamefont {Grissonnanche}, \citenamefont {Baglo},
  \citenamefont {Lampen-Kelley}, \citenamefont {Yan}, \citenamefont {Balz},
  \citenamefont {Mandrus}, \citenamefont {Nagler}, \citenamefont {Kim},
  \citenamefont {Kim}, \citenamefont {Doiron-Leyraud},\ and\ \citenamefont
  {Taillefer}}]{Lefrancois2022}%
  \BibitemOpen
  \bibfield  {author} {\bibinfo {author} {\bibfnamefont {E.}~\bibnamefont
  {Lefran\ifmmode~\mbox{\c{c}}\else \c{c}\fi{}ois}}, \bibinfo {author}
  {\bibfnamefont {G.}~\bibnamefont {Grissonnanche}}, \bibinfo {author}
  {\bibfnamefont {J.}~\bibnamefont {Baglo}}, \bibinfo {author} {\bibfnamefont
  {P.}~\bibnamefont {Lampen-Kelley}}, \bibinfo {author} {\bibfnamefont {J.-Q.}\
  \bibnamefont {Yan}}, \bibinfo {author} {\bibfnamefont {C.}~\bibnamefont
  {Balz}}, \bibinfo {author} {\bibfnamefont {D.}~\bibnamefont {Mandrus}},
  \bibinfo {author} {\bibfnamefont {S.~E.}\ \bibnamefont {Nagler}}, \bibinfo
  {author} {\bibfnamefont {S.}~\bibnamefont {Kim}}, \bibinfo {author}
  {\bibfnamefont {Y.-J.}\ \bibnamefont {Kim}}, \bibinfo {author} {\bibfnamefont
  {N.}~\bibnamefont {Doiron-Leyraud}},\ and\ \bibinfo {author} {\bibfnamefont
  {L.}~\bibnamefont {Taillefer}},\ }\bibfield  {title} {\bibinfo {title}
  {Evidence of a phonon hall effect in the kitaev spin liquid candidate
  ${\alpha}$-{RuCl}$_3$},\ }\href {https://doi.org/10.1103/PhysRevX.12.021025}
  {\bibfield  {journal} {\bibinfo  {journal} {Phys. Rev. X}\ }\textbf {\bibinfo
  {volume} {12}},\ \bibinfo {pages} {021025} (\bibinfo {year}
  {2022})}\BibitemShut {NoStop}%
\bibitem [{\citenamefont {Winter}\ \emph
  {et~al.}(2017{\natexlab{a}})\citenamefont {Winter}, \citenamefont {Tsirlin},
  \citenamefont {Daghofer}, \citenamefont {Brink}, \citenamefont {Singh},
  \citenamefont {Gegenwart},\ and\ \citenamefont {Valentí}}]{Winter2017}%
  \BibitemOpen
  \bibfield  {author} {\bibinfo {author} {\bibfnamefont {S.~M.}\ \bibnamefont
  {Winter}}, \bibinfo {author} {\bibfnamefont {A.~A.}\ \bibnamefont {Tsirlin}},
  \bibinfo {author} {\bibfnamefont {M.}~\bibnamefont {Daghofer}}, \bibinfo
  {author} {\bibfnamefont {J.~v.~d.}\ \bibnamefont {Brink}}, \bibinfo {author}
  {\bibfnamefont {Y.}~\bibnamefont {Singh}}, \bibinfo {author} {\bibfnamefont
  {P.}~\bibnamefont {Gegenwart}},\ and\ \bibinfo {author} {\bibfnamefont
  {R.}~\bibnamefont {Valentí}},\ }\bibfield  {title} {\bibinfo {title} {Models
  and materials for generalized kitaev magnetism},\ }\href
  {https://doi.org/10.1088/1361-648X/aa8cf5} {\bibfield  {journal} {\bibinfo
  {journal} {J. Phys.: Condens. Matter}\ }\textbf {\bibinfo {volume} {29}},\
  \bibinfo {pages} {493002} (\bibinfo {year} {2017}{\natexlab{a}})}\BibitemShut
  {NoStop}%
\bibitem [{\citenamefont {Takagi}\ \emph {et~al.}(2019)\citenamefont {Takagi},
  \citenamefont {Takayama}, \citenamefont {Jackeli}, \citenamefont
  {Khaliullin},\ and\ \citenamefont {Nagler}}]{Takagi2019}%
  \BibitemOpen
  \bibfield  {author} {\bibinfo {author} {\bibfnamefont {H.}~\bibnamefont
  {Takagi}}, \bibinfo {author} {\bibfnamefont {T.}~\bibnamefont {Takayama}},
  \bibinfo {author} {\bibfnamefont {G.}~\bibnamefont {Jackeli}}, \bibinfo
  {author} {\bibfnamefont {G.}~\bibnamefont {Khaliullin}},\ and\ \bibinfo
  {author} {\bibfnamefont {S.~E.}\ \bibnamefont {Nagler}},\ }\bibfield  {title}
  {\bibinfo {title} {Concept and realization of {Kitaev} quantum spin
  liquids},\ }\href {https://doi.org/10.1038/s42254-019-0038-2} {\bibfield
  {journal} {\bibinfo  {journal} {Nat. Rev. Phys.}\ }\textbf {\bibinfo {volume}
  {1}},\ \bibinfo {pages} {264} (\bibinfo {year} {2019})}\BibitemShut {NoStop}%
\bibitem [{\citenamefont {Motome}\ and\ \citenamefont
  {Nasu}(2020)}]{Motome2020review}%
  \BibitemOpen
  \bibfield  {author} {\bibinfo {author} {\bibfnamefont {Y.}~\bibnamefont
  {Motome}}\ and\ \bibinfo {author} {\bibfnamefont {J.}~\bibnamefont {Nasu}},\
  }\bibfield  {title} {\bibinfo {title} {Hunting {Majorana} {Fermions} in
  {Kitaev} {Magnets}},\ }\href {https://doi.org/10.7566/JPSJ.89.012002}
  {\bibfield  {journal} {\bibinfo  {journal} {J. Phys. Soc. Jpn.}\ }\textbf
  {\bibinfo {volume} {89}},\ \bibinfo {pages} {012002} (\bibinfo {year}
  {2020})}\BibitemShut {NoStop}%
\bibitem [{\citenamefont {Kim}\ \emph {et~al.}(2022)\citenamefont {Kim},
  \citenamefont {Yuan},\ and\ \citenamefont {Kim}}]{Kim2022}%
  \BibitemOpen
  \bibfield  {author} {\bibinfo {author} {\bibfnamefont {S.}~\bibnamefont
  {Kim}}, \bibinfo {author} {\bibfnamefont {B.}~\bibnamefont {Yuan}},\ and\
  \bibinfo {author} {\bibfnamefont {Y.-J.}\ \bibnamefont {Kim}},\ }\bibfield
  {title} {\bibinfo {title} {{$\alpha$-RuCl$_3$ and other Kitaev materials}},\
  }\bibfield  {journal} {\bibinfo  {journal} {APL Materials}\ }\textbf
  {\bibinfo {volume} {10}},\ \href {https://doi.org/10.1063/5.0101512}
  {10.1063/5.0101512} (\bibinfo {year} {2022}),\ \bibinfo {note} {080903},\
  \Eprint
  {https://arxiv.org/abs/https://pubs.aip.org/aip/apm/article-pdf/doi/10.1063/5.0101512/16490529/080903\_1\_online.pdf}
  {https://pubs.aip.org/aip/apm/article-pdf/doi/10.1063/5.0101512/16490529/080903\_1\_online.pdf}
  \BibitemShut {NoStop}%
\bibitem [{\citenamefont {Kitaev}(2006)}]{kitaev2006}%
  \BibitemOpen
  \bibfield  {author} {\bibinfo {author} {\bibfnamefont {A.}~\bibnamefont
  {Kitaev}},\ }\bibfield  {title} {\bibinfo {title} {Anyons in an exactly
  solved model and beyond},\ }\href
  {https://doi.org/http://dx.doi.org/10.1016/j.aop.2005.10.005} {\bibfield
  {journal} {\bibinfo  {journal} {Annals of Physics}\ }\textbf {\bibinfo
  {volume} {321}},\ \bibinfo {pages} {2 } (\bibinfo {year} {2006})},\ \bibinfo
  {note} {january Special Issue}\BibitemShut {NoStop}%
\bibitem [{\citenamefont {Hermanns}\ \emph {et~al.}(2018)\citenamefont
  {Hermanns}, \citenamefont {Kimchi},\ and\ \citenamefont
  {Knolle}}]{Hermanns2018}%
  \BibitemOpen
  \bibfield  {author} {\bibinfo {author} {\bibfnamefont {M.}~\bibnamefont
  {Hermanns}}, \bibinfo {author} {\bibfnamefont {I.}~\bibnamefont {Kimchi}},\
  and\ \bibinfo {author} {\bibfnamefont {J.}~\bibnamefont {Knolle}},\
  }\bibfield  {title} {\bibinfo {title} {Physics of the kitaev model:
  Fractionalization, dynamic correlations, and material connections},\ }\href
  {https://doi.org/10.1146/annurev-conmatphys-033117-053934} {\bibfield
  {journal} {\bibinfo  {journal} {Annual Review of Condensed Matter Physics}\
  }\textbf {\bibinfo {volume} {9}},\ \bibinfo {pages} {17} (\bibinfo {year}
  {2018})},\ \Eprint
  {https://arxiv.org/abs/https://doi.org/10.1146/annurev-conmatphys-033117-053934}
  {https://doi.org/10.1146/annurev-conmatphys-033117-053934} \BibitemShut
  {NoStop}%
\bibitem [{\citenamefont {Czajka}\ \emph {et~al.}(2021)\citenamefont {Czajka},
  \citenamefont {Gao}, \citenamefont {Hirschberger}, \citenamefont
  {Lampen-Kelley}, \citenamefont {Banerjee}, \citenamefont {Yan}, \citenamefont
  {Mandrus}, \citenamefont {Nagler},\ and\ \citenamefont {Ong}}]{czajka2021}%
  \BibitemOpen
  \bibfield  {author} {\bibinfo {author} {\bibfnamefont {P.}~\bibnamefont
  {Czajka}}, \bibinfo {author} {\bibfnamefont {T.}~\bibnamefont {Gao}},
  \bibinfo {author} {\bibfnamefont {M.}~\bibnamefont {Hirschberger}}, \bibinfo
  {author} {\bibfnamefont {P.}~\bibnamefont {Lampen-Kelley}}, \bibinfo {author}
  {\bibfnamefont {A.}~\bibnamefont {Banerjee}}, \bibinfo {author}
  {\bibfnamefont {J.}~\bibnamefont {Yan}}, \bibinfo {author} {\bibfnamefont
  {D.~G.}\ \bibnamefont {Mandrus}}, \bibinfo {author} {\bibfnamefont {S.~E.}\
  \bibnamefont {Nagler}},\ and\ \bibinfo {author} {\bibfnamefont {N.~P.}\
  \bibnamefont {Ong}},\ }\bibfield  {title} {\bibinfo {title} {Oscillations of
  the thermal conductivity in the spin-liquid state of $\alpha$-{RuCl}$_3$},\
  }\href {https://doi.org/10.1038/s41567-021-01243-x} {\bibfield  {journal}
  {\bibinfo  {journal} {Nat. Phys.}\ }\textbf {\bibinfo {volume} {17}},\
  \bibinfo {pages} {915} (\bibinfo {year} {2021})}\BibitemShut {NoStop}%
\bibitem [{\citenamefont {Rau}\ \emph {et~al.}(2014)\citenamefont {Rau},
  \citenamefont {Lee},\ and\ \citenamefont {Kee}}]{Rau2014}%
  \BibitemOpen
  \bibfield  {author} {\bibinfo {author} {\bibfnamefont {J.~G.}\ \bibnamefont
  {Rau}}, \bibinfo {author} {\bibfnamefont {E.~K.-H.}\ \bibnamefont {Lee}},\
  and\ \bibinfo {author} {\bibfnamefont {H.-Y.}\ \bibnamefont {Kee}},\
  }\bibfield  {title} {\bibinfo {title} {Generic spin model for the honeycomb
  iridates beyond the kitaev limit},\ }\href
  {https://doi.org/10.1103/PhysRevLett.112.077204} {\bibfield  {journal}
  {\bibinfo  {journal} {Phys. Rev. Lett.}\ }\textbf {\bibinfo {volume} {112}},\
  \bibinfo {pages} {077204} (\bibinfo {year} {2014})}\BibitemShut {NoStop}%
\bibitem [{\citenamefont {Winter}\ \emph
  {et~al.}(2017{\natexlab{b}})\citenamefont {Winter}, \citenamefont {Riedl},
  \citenamefont {Maksimov}, \citenamefont {Chernyshev}, \citenamefont
  {Honecker},\ and\ \citenamefont {Valent\'{\i}}}]{winter2017NC}%
  \BibitemOpen
  \bibfield  {author} {\bibinfo {author} {\bibfnamefont {S.~M.}\ \bibnamefont
  {Winter}}, \bibinfo {author} {\bibfnamefont {K.}~\bibnamefont {Riedl}},
  \bibinfo {author} {\bibfnamefont {P.~A.}\ \bibnamefont {Maksimov}}, \bibinfo
  {author} {\bibfnamefont {A.~L.}\ \bibnamefont {Chernyshev}}, \bibinfo
  {author} {\bibfnamefont {A.}~\bibnamefont {Honecker}},\ and\ \bibinfo
  {author} {\bibfnamefont {R.}~\bibnamefont {Valent\'{\i}}},\ }\bibfield
  {title} {\bibinfo {title} {Breakdown of magnons in a strongly spin-orbital
  coupled magnet},\ }\href {https://doi.org/10.1038/s41467-017-01177-0}
  {\bibfield  {journal} {\bibinfo  {journal} {Nature Communications}\ }\textbf
  {\bibinfo {volume} {8}},\ \bibinfo {pages} {1152} (\bibinfo {year}
  {2017}{\natexlab{b}})}\BibitemShut {NoStop}%
\bibitem [{\citenamefont {Janssen}\ \emph {et~al.}(2017)\citenamefont
  {Janssen}, \citenamefont {Andrade},\ and\ \citenamefont
  {Vojta}}]{janssen2017}%
  \BibitemOpen
  \bibfield  {author} {\bibinfo {author} {\bibfnamefont {L.}~\bibnamefont
  {Janssen}}, \bibinfo {author} {\bibfnamefont {E.~C.}\ \bibnamefont
  {Andrade}},\ and\ \bibinfo {author} {\bibfnamefont {M.}~\bibnamefont
  {Vojta}},\ }\bibfield  {title} {\bibinfo {title} {Magnetization processes of
  zigzag states on the honeycomb lattice: {Identifying} spin models for
  $\alpha$-{RuCl}$_3$ and {Na}$_2${IrO}$_3$},\ }\href
  {https://doi.org/10.1103/PhysRevB.96.064430} {\bibfield  {journal} {\bibinfo
  {journal} {Phys. Rev. B}\ }\textbf {\bibinfo {volume} {96}},\ \bibinfo
  {pages} {064430} (\bibinfo {year} {2017})}\BibitemShut {NoStop}%
\bibitem [{\citenamefont {Modic}\ \emph
  {et~al.}(2018{\natexlab{b}})\citenamefont {Modic}, \citenamefont {Ramshaw},
  \citenamefont {Shekhter},\ and\ \citenamefont {Varma}}]{modic2018}%
  \BibitemOpen
  \bibfield  {author} {\bibinfo {author} {\bibfnamefont {K.~A.}\ \bibnamefont
  {Modic}}, \bibinfo {author} {\bibfnamefont {B.~J.}\ \bibnamefont {Ramshaw}},
  \bibinfo {author} {\bibfnamefont {A.}~\bibnamefont {Shekhter}},\ and\
  \bibinfo {author} {\bibfnamefont {C.~M.}\ \bibnamefont {Varma}},\ }\bibfield
  {title} {\bibinfo {title} {Chiral spin order in some purported {Kitaev}
  spin-liquid compounds},\ }\href {https://doi.org/10.1103/PhysRevB.98.205110}
  {\bibfield  {journal} {\bibinfo  {journal} {Phys. Rev. B}\ }\textbf {\bibinfo
  {volume} {98}},\ \bibinfo {pages} {205110} (\bibinfo {year}
  {2018}{\natexlab{b}})}\BibitemShut {NoStop}%
\bibitem [{\citenamefont {Ran}\ \emph {et~al.}(2017)\citenamefont {Ran},
  \citenamefont {Wang}, \citenamefont {Wang}, \citenamefont {Dong},
  \citenamefont {Ren}, \citenamefont {Bao}, \citenamefont {Li}, \citenamefont
  {Ma}, \citenamefont {Gan}, \citenamefont {Zhang}, \citenamefont {Park},
  \citenamefont {Deng}, \citenamefont {Danilkin}, \citenamefont {Yu},
  \citenamefont {Li},\ and\ \citenamefont {Wen}}]{ran2017}%
  \BibitemOpen
  \bibfield  {author} {\bibinfo {author} {\bibfnamefont {K.}~\bibnamefont
  {Ran}}, \bibinfo {author} {\bibfnamefont {J.}~\bibnamefont {Wang}}, \bibinfo
  {author} {\bibfnamefont {W.}~\bibnamefont {Wang}}, \bibinfo {author}
  {\bibfnamefont {Z.-Y.}\ \bibnamefont {Dong}}, \bibinfo {author}
  {\bibfnamefont {X.}~\bibnamefont {Ren}}, \bibinfo {author} {\bibfnamefont
  {S.}~\bibnamefont {Bao}}, \bibinfo {author} {\bibfnamefont {S.}~\bibnamefont
  {Li}}, \bibinfo {author} {\bibfnamefont {Z.}~\bibnamefont {Ma}}, \bibinfo
  {author} {\bibfnamefont {Y.}~\bibnamefont {Gan}}, \bibinfo {author}
  {\bibfnamefont {Y.}~\bibnamefont {Zhang}}, \bibinfo {author} {\bibfnamefont
  {J.}~\bibnamefont {Park}}, \bibinfo {author} {\bibfnamefont {G.}~\bibnamefont
  {Deng}}, \bibinfo {author} {\bibfnamefont {S.}~\bibnamefont {Danilkin}},
  \bibinfo {author} {\bibfnamefont {S.-L.}\ \bibnamefont {Yu}}, \bibinfo
  {author} {\bibfnamefont {J.-X.}\ \bibnamefont {Li}},\ and\ \bibinfo {author}
  {\bibfnamefont {J.}~\bibnamefont {Wen}},\ }\bibfield  {title} {\bibinfo
  {title} {Spin-{Wave} {Excitations} {Evidencing} the {Kitaev} {Interaction} in
  {Single} {Crystalline} $\alpha$-{RuCl}$_3$},\ }\href
  {https://doi.org/10.1103/PhysRevLett.118.107203} {\bibfield  {journal}
  {\bibinfo  {journal} {Phys. Rev. Lett.}\ }\textbf {\bibinfo {volume} {118}},\
  \bibinfo {pages} {107203} (\bibinfo {year} {2017})}\BibitemShut {NoStop}%
\bibitem [{\citenamefont {Riedl}\ \emph {et~al.}(2019)\citenamefont {Riedl},
  \citenamefont {Li}, \citenamefont {Winter},\ and\ \citenamefont
  {Valentí}}]{riedl2019}%
  \BibitemOpen
  \bibfield  {author} {\bibinfo {author} {\bibfnamefont {K.}~\bibnamefont
  {Riedl}}, \bibinfo {author} {\bibfnamefont {Y.}~\bibnamefont {Li}}, \bibinfo
  {author} {\bibfnamefont {S.~M.}\ \bibnamefont {Winter}},\ and\ \bibinfo
  {author} {\bibfnamefont {R.}~\bibnamefont {Valentí}},\ }\bibfield  {title}
  {\bibinfo {title} {Sawtooth {Torque} in {Anisotropic} $j_{eff}=1/2$
  {Magnets}: {Application} to $\alpha$-{RuCl}$_3$},\ }\href
  {https://doi.org/10.1103/PhysRevLett.122.197202} {\bibfield  {journal}
  {\bibinfo  {journal} {Phys. Rev. Lett.}\ }\textbf {\bibinfo {volume} {122}},\
  \bibinfo {pages} {197202} (\bibinfo {year} {2019})}\BibitemShut {NoStop}%
\bibitem [{\citenamefont {Koitzsch}\ \emph {et~al.}(2020)\citenamefont
  {Koitzsch}, \citenamefont {Müller}, \citenamefont {Knupfer}, \citenamefont
  {Büchner}, \citenamefont {Nowak}, \citenamefont {Isaeva}, \citenamefont
  {Doert}, \citenamefont {Grüninger}, \citenamefont {Nishimoto},\ and\
  \citenamefont {van~den Brink}}]{koitzsch2020}%
  \BibitemOpen
  \bibfield  {author} {\bibinfo {author} {\bibfnamefont {A.}~\bibnamefont
  {Koitzsch}}, \bibinfo {author} {\bibfnamefont {E.}~\bibnamefont {Müller}},
  \bibinfo {author} {\bibfnamefont {M.}~\bibnamefont {Knupfer}}, \bibinfo
  {author} {\bibfnamefont {B.}~\bibnamefont {Büchner}}, \bibinfo {author}
  {\bibfnamefont {D.}~\bibnamefont {Nowak}}, \bibinfo {author} {\bibfnamefont
  {A.}~\bibnamefont {Isaeva}}, \bibinfo {author} {\bibfnamefont
  {T.}~\bibnamefont {Doert}}, \bibinfo {author} {\bibfnamefont
  {M.}~\bibnamefont {Grüninger}}, \bibinfo {author} {\bibfnamefont
  {S.}~\bibnamefont {Nishimoto}},\ and\ \bibinfo {author} {\bibfnamefont
  {J.}~\bibnamefont {van~den Brink}},\ }\bibfield  {title} {\bibinfo {title}
  {Low-temperature enhancement of ferromagnetic {Kitaev} correlations in
  $\alpha$-{RuCl}$_3$},\ }\href
  {https://doi.org/10.1103/PhysRevMaterials.4.094408} {\bibfield  {journal}
  {\bibinfo  {journal} {Phys. Rev. Materials}\ }\textbf {\bibinfo {volume}
  {4}},\ \bibinfo {pages} {094408} (\bibinfo {year} {2020})}\BibitemShut
  {NoStop}%
\bibitem [{\citenamefont {Sears}\ \emph {et~al.}(2020)\citenamefont {Sears},
  \citenamefont {Chern}, \citenamefont {Kim}, \citenamefont {Bereciartua},
  \citenamefont {Francoual}, \citenamefont {Kim},\ and\ \citenamefont
  {Kim}}]{Sears2020}%
  \BibitemOpen
  \bibfield  {author} {\bibinfo {author} {\bibfnamefont {J.~A.}\ \bibnamefont
  {Sears}}, \bibinfo {author} {\bibfnamefont {L.~E.}\ \bibnamefont {Chern}},
  \bibinfo {author} {\bibfnamefont {S.}~\bibnamefont {Kim}}, \bibinfo {author}
  {\bibfnamefont {P.~J.}\ \bibnamefont {Bereciartua}}, \bibinfo {author}
  {\bibfnamefont {S.}~\bibnamefont {Francoual}}, \bibinfo {author}
  {\bibfnamefont {Y.~B.}\ \bibnamefont {Kim}},\ and\ \bibinfo {author}
  {\bibfnamefont {Y.-J.}\ \bibnamefont {Kim}},\ }\bibfield  {title} {\bibinfo
  {title} {Ferromagnetic kitaev interaction and the origin of large magnetic
  anisotropy in ${\alpha}$-{RuCl}$_{3}$},\ }\href
  {https://doi.org/10.1038/s41567-020-0874-0} {\bibfield  {journal} {\bibinfo
  {journal} {Nature Physics}\ }\textbf {\bibinfo {volume} {16}},\ \bibinfo
  {pages} {837} (\bibinfo {year} {2020})}\BibitemShut {NoStop}%
\bibitem [{\citenamefont {Lee}\ \emph {et~al.}(2016)\citenamefont {Lee},
  \citenamefont {Rau},\ and\ \citenamefont {Kim}}]{Lee2016}%
  \BibitemOpen
  \bibfield  {author} {\bibinfo {author} {\bibfnamefont {E.~K.-H.}\
  \bibnamefont {Lee}}, \bibinfo {author} {\bibfnamefont {J.~G.}\ \bibnamefont
  {Rau}},\ and\ \bibinfo {author} {\bibfnamefont {Y.~B.}\ \bibnamefont {Kim}},\
  }\bibfield  {title} {\bibinfo {title} {Two iridates, two models, and two
  approaches: A comparative study on magnetism in three-dimensional honeycomb
  materials},\ }\href {https://doi.org/10.1103/PhysRevB.93.184420} {\bibfield
  {journal} {\bibinfo  {journal} {Phys. Rev. B}\ }\textbf {\bibinfo {volume}
  {93}},\ \bibinfo {pages} {184420} (\bibinfo {year} {2016})}\BibitemShut
  {NoStop}%
\bibitem [{\citenamefont {Kimchi}\ \emph {et~al.}(2015)\citenamefont {Kimchi},
  \citenamefont {Coldea},\ and\ \citenamefont {Vishwanath}}]{Kimchi2015}%
  \BibitemOpen
  \bibfield  {author} {\bibinfo {author} {\bibfnamefont {I.}~\bibnamefont
  {Kimchi}}, \bibinfo {author} {\bibfnamefont {R.}~\bibnamefont {Coldea}},\
  and\ \bibinfo {author} {\bibfnamefont {A.}~\bibnamefont {Vishwanath}},\
  }\bibfield  {title} {\bibinfo {title} {Unified theory of spiral magnetism in
  the harmonic-honeycomb iridates $\ensuremath{\alpha},\ensuremath{\beta}$, and
  $\ensuremath{\gamma}\phantom{\rule{0.16em}{0ex}}{\mathrm{li}}_{2}{\mathrm{iro}}_{3}$},\
  }\href {https://doi.org/10.1103/PhysRevB.91.245134} {\bibfield  {journal}
  {\bibinfo  {journal} {Phys. Rev. B}\ }\textbf {\bibinfo {volume} {91}},\
  \bibinfo {pages} {245134} (\bibinfo {year} {2015})}\BibitemShut {NoStop}%
\bibitem [{\citenamefont {Winter}\ \emph {et~al.}(2016)\citenamefont {Winter},
  \citenamefont {Li}, \citenamefont {Jeschke},\ and\ \citenamefont
  {Valent\'{\i}}}]{Winter2016}%
  \BibitemOpen
  \bibfield  {author} {\bibinfo {author} {\bibfnamefont {S.~M.}\ \bibnamefont
  {Winter}}, \bibinfo {author} {\bibfnamefont {Y.}~\bibnamefont {Li}}, \bibinfo
  {author} {\bibfnamefont {H.~O.}\ \bibnamefont {Jeschke}},\ and\ \bibinfo
  {author} {\bibfnamefont {R.}~\bibnamefont {Valent\'{\i}}},\ }\bibfield
  {title} {\bibinfo {title} {Challenges in design of kitaev materials: Magnetic
  interactions from competing energy scales},\ }\href
  {https://doi.org/10.1103/PhysRevB.93.214431} {\bibfield  {journal} {\bibinfo
  {journal} {Phys. Rev. B}\ }\textbf {\bibinfo {volume} {93}},\ \bibinfo
  {pages} {214431} (\bibinfo {year} {2016})}\BibitemShut {NoStop}%
\bibitem [{\citenamefont {Gordon}\ \emph {et~al.}(2019)\citenamefont {Gordon},
  \citenamefont {Catuneanu}, \citenamefont {Sørensen},\ and\ \citenamefont
  {Kee}}]{gordon2019}%
  \BibitemOpen
  \bibfield  {author} {\bibinfo {author} {\bibfnamefont {J.~S.}\ \bibnamefont
  {Gordon}}, \bibinfo {author} {\bibfnamefont {A.}~\bibnamefont {Catuneanu}},
  \bibinfo {author} {\bibfnamefont {E.~S.}\ \bibnamefont {Sørensen}},\ and\
  \bibinfo {author} {\bibfnamefont {H.-Y.}\ \bibnamefont {Kee}},\ }\bibfield
  {title} {\bibinfo {title} {Theory of the field-revealed {Kitaev} spin
  liquid},\ }\href {https://doi.org/10.1038/s41467-019-10405-8} {\bibfield
  {journal} {\bibinfo  {journal} {Nat Commun}\ }\textbf {\bibinfo {volume}
  {10}},\ \bibinfo {pages} {2470} (\bibinfo {year} {2019})}\BibitemShut
  {NoStop}%
\bibitem [{\citenamefont {{Park}}\ \emph {et~al.}(2016)\citenamefont {{Park}},
  \citenamefont {{Do}}, \citenamefont {{Choi}}, \citenamefont {{Jang}},
  \citenamefont {{Jang}}, \citenamefont {{Schefer}}, \citenamefont {{Wu}},
  \citenamefont {{Gardner}}, \citenamefont {{Park}}, \citenamefont {{Park}},\
  and\ \citenamefont {{Ji}}}]{park2016}%
  \BibitemOpen
  \bibfield  {author} {\bibinfo {author} {\bibfnamefont {S.-Y.}\ \bibnamefont
  {{Park}}}, \bibinfo {author} {\bibfnamefont {S.-H.}\ \bibnamefont {{Do}}},
  \bibinfo {author} {\bibfnamefont {K.-Y.}\ \bibnamefont {{Choi}}}, \bibinfo
  {author} {\bibfnamefont {D.}~\bibnamefont {{Jang}}}, \bibinfo {author}
  {\bibfnamefont {T.-H.}\ \bibnamefont {{Jang}}}, \bibinfo {author}
  {\bibfnamefont {J.}~\bibnamefont {{Schefer}}}, \bibinfo {author}
  {\bibfnamefont {C.-M.}\ \bibnamefont {{Wu}}}, \bibinfo {author}
  {\bibfnamefont {J.~S.}\ \bibnamefont {{Gardner}}}, \bibinfo {author}
  {\bibfnamefont {J.~M.~S.}\ \bibnamefont {{Park}}}, \bibinfo {author}
  {\bibfnamefont {J.-H.}\ \bibnamefont {{Park}}},\ and\ \bibinfo {author}
  {\bibfnamefont {S.}~\bibnamefont {{Ji}}},\ }\bibfield  {title} {\bibinfo
  {title} {{Emergence of the Isotropic Kitaev Honeycomb Lattice with
  Two-dimensional Ising Universality in $\alpha$-RuCl$_3$}},\ }\href@noop {}
  {\bibfield  {journal} {\bibinfo  {journal} {ArXiv e-prints}\ } (\bibinfo
  {year} {2016})},\ \Eprint {https://arxiv.org/abs/1609.05690}
  {arXiv:1609.05690 [cond-mat.mtrl-sci]} \BibitemShut {NoStop}%
\bibitem [{\citenamefont {Kasahara}\ \emph {et~al.}(2022)\citenamefont
  {Kasahara}, \citenamefont {Suetsugu}, \citenamefont {Asaba}, \citenamefont
  {Kasahara}, \citenamefont {Shibauchi}, \citenamefont {Kurita}, \citenamefont
  {Tanaka},\ and\ \citenamefont {Matsuda}}]{Kasahara2022}%
  \BibitemOpen
  \bibfield  {author} {\bibinfo {author} {\bibfnamefont {Y.}~\bibnamefont
  {Kasahara}}, \bibinfo {author} {\bibfnamefont {S.}~\bibnamefont {Suetsugu}},
  \bibinfo {author} {\bibfnamefont {T.}~\bibnamefont {Asaba}}, \bibinfo
  {author} {\bibfnamefont {S.}~\bibnamefont {Kasahara}}, \bibinfo {author}
  {\bibfnamefont {T.}~\bibnamefont {Shibauchi}}, \bibinfo {author}
  {\bibfnamefont {N.}~\bibnamefont {Kurita}}, \bibinfo {author} {\bibfnamefont
  {H.}~\bibnamefont {Tanaka}},\ and\ \bibinfo {author} {\bibfnamefont
  {Y.}~\bibnamefont {Matsuda}},\ }\href
  {https://doi.org/10.48550/ARXIV.2202.11947} {\bibinfo {title} {Quantized and
  unquantized thermal hall conductance of kitaev spin-liquid candidate
  $\alpha$-{RuCl}$_3$}} (\bibinfo {year} {2022}),\ \Eprint
  {https://arxiv.org/abs/2202.11947} {arXiv:2202.11947} \BibitemShut {NoStop}%
\bibitem [{\citenamefont {Zhang}\ \emph {et~al.}(2023)\citenamefont {Zhang},
  \citenamefont {McGuire}, \citenamefont {May}, \citenamefont {Chao},
  \citenamefont {Zheng}, \citenamefont {Chi}, \citenamefont {Sales},
  \citenamefont {Mandrus}, \citenamefont {Nagler}, \citenamefont {Miao},
  \citenamefont {Ye},\ and\ \citenamefont {Yan}}]{Zhang2023}%
  \BibitemOpen
  \bibfield  {author} {\bibinfo {author} {\bibfnamefont {H.}~\bibnamefont
  {Zhang}}, \bibinfo {author} {\bibfnamefont {M.}~\bibnamefont {McGuire}},
  \bibinfo {author} {\bibfnamefont {A.~F.}\ \bibnamefont {May}}, \bibinfo
  {author} {\bibfnamefont {J.}~\bibnamefont {Chao}}, \bibinfo {author}
  {\bibfnamefont {Q.}~\bibnamefont {Zheng}}, \bibinfo {author} {\bibfnamefont
  {M.}~\bibnamefont {Chi}}, \bibinfo {author} {\bibfnamefont {B.~C.}\
  \bibnamefont {Sales}}, \bibinfo {author} {\bibfnamefont {D.~G.}\ \bibnamefont
  {Mandrus}}, \bibinfo {author} {\bibfnamefont {S.~E.}\ \bibnamefont {Nagler}},
  \bibinfo {author} {\bibfnamefont {H.}~\bibnamefont {Miao}}, \bibinfo {author}
  {\bibfnamefont {F.}~\bibnamefont {Ye}},\ and\ \bibinfo {author}
  {\bibfnamefont {J.}~\bibnamefont {Yan}},\ }\bibfield  {title} {\bibinfo
  {title} {{Stacking disorder and thermal transport properties of
  $\alpha$-RuCl$_3$}},\ }\href@noop {} {\bibfield  {journal} {\bibinfo
  {journal} {ArXiv e-prints}\ } (\bibinfo {year} {2023})},\ \Eprint
  {https://arxiv.org/abs/2303.03682} {arXiv:2303.03682 [cond-mat.mtrl-sci]}
  \BibitemShut {NoStop}%
\bibitem [{\citenamefont {Glamazda}\ \emph {et~al.}(2017)\citenamefont
  {Glamazda}, \citenamefont {Lemmens}, \citenamefont {Do}, \citenamefont
  {Kwon},\ and\ \citenamefont {Choi}}]{Glamazda2017}%
  \BibitemOpen
  \bibfield  {author} {\bibinfo {author} {\bibfnamefont {A.}~\bibnamefont
  {Glamazda}}, \bibinfo {author} {\bibfnamefont {P.}~\bibnamefont {Lemmens}},
  \bibinfo {author} {\bibfnamefont {S.-H.}\ \bibnamefont {Do}}, \bibinfo
  {author} {\bibfnamefont {Y.~S.}\ \bibnamefont {Kwon}},\ and\ \bibinfo
  {author} {\bibfnamefont {K.-Y.}\ \bibnamefont {Choi}},\ }\bibfield  {title}
  {\bibinfo {title} {Relation between kitaev magnetism and structure in
  $\ensuremath{\alpha}\ensuremath{-}{\mathrm{rucl}}_{3}$},\ }\href
  {https://doi.org/10.1103/PhysRevB.95.174429} {\bibfield  {journal} {\bibinfo
  {journal} {Phys. Rev. B}\ }\textbf {\bibinfo {volume} {95}},\ \bibinfo
  {pages} {174429} (\bibinfo {year} {2017})}\BibitemShut {NoStop}%
\bibitem [{\citenamefont {Mu}\ \emph {et~al.}(2022)\citenamefont {Mu},
  \citenamefont {Dixit}, \citenamefont {Wang}, \citenamefont {Abernathy},
  \citenamefont {Cao}, \citenamefont {Nagler}, \citenamefont {Yan},
  \citenamefont {Lampen-Kelley}, \citenamefont {Mandrus}, \citenamefont
  {Polanco}, \citenamefont {Liang}, \citenamefont {Hal\'asz}, \citenamefont
  {Cheng}, \citenamefont {Banerjee},\ and\ \citenamefont {Berlijn}}]{Mu2022}%
  \BibitemOpen
  \bibfield  {author} {\bibinfo {author} {\bibfnamefont {S.}~\bibnamefont
  {Mu}}, \bibinfo {author} {\bibfnamefont {K.~D.}\ \bibnamefont {Dixit}},
  \bibinfo {author} {\bibfnamefont {X.}~\bibnamefont {Wang}}, \bibinfo {author}
  {\bibfnamefont {D.~L.}\ \bibnamefont {Abernathy}}, \bibinfo {author}
  {\bibfnamefont {H.}~\bibnamefont {Cao}}, \bibinfo {author} {\bibfnamefont
  {S.~E.}\ \bibnamefont {Nagler}}, \bibinfo {author} {\bibfnamefont
  {J.}~\bibnamefont {Yan}}, \bibinfo {author} {\bibfnamefont {P.}~\bibnamefont
  {Lampen-Kelley}}, \bibinfo {author} {\bibfnamefont {D.}~\bibnamefont
  {Mandrus}}, \bibinfo {author} {\bibfnamefont {C.~A.}\ \bibnamefont
  {Polanco}}, \bibinfo {author} {\bibfnamefont {L.}~\bibnamefont {Liang}},
  \bibinfo {author} {\bibfnamefont {G.~B.}\ \bibnamefont {Hal\'asz}}, \bibinfo
  {author} {\bibfnamefont {Y.}~\bibnamefont {Cheng}}, \bibinfo {author}
  {\bibfnamefont {A.}~\bibnamefont {Banerjee}},\ and\ \bibinfo {author}
  {\bibfnamefont {T.}~\bibnamefont {Berlijn}},\ }\bibfield  {title} {\bibinfo
  {title} {Role of the third dimension in searching for majorana fermions in
  $\ensuremath{\alpha}\text{\ensuremath{-}}{\mathrm{rucl}}_{3}$ via phonons},\
  }\href {https://doi.org/10.1103/PhysRevResearch.4.013067} {\bibfield
  {journal} {\bibinfo  {journal} {Phys. Rev. Res.}\ }\textbf {\bibinfo {volume}
  {4}},\ \bibinfo {pages} {013067} (\bibinfo {year} {2022})}\BibitemShut
  {NoStop}%
\bibitem [{\citenamefont {Yang}\ \emph {et~al.}(2023)\citenamefont {Yang},
  \citenamefont {Goh}, \citenamefont {Sung}, \citenamefont {Ye}, \citenamefont
  {Biswas}, \citenamefont {Kaib}, \citenamefont {Dhakal}, \citenamefont {Yan},
  \citenamefont {Li}, \citenamefont {Jiang}, \citenamefont {Chen},
  \citenamefont {Lei}, \citenamefont {He}, \citenamefont {Valentí},
  \citenamefont {Winter}, \citenamefont {Hovden},\ and\ \citenamefont
  {Tsen}}]{yang_magnetic_2023}%
  \BibitemOpen
  \bibfield  {author} {\bibinfo {author} {\bibfnamefont {B.}~\bibnamefont
  {Yang}}, \bibinfo {author} {\bibfnamefont {Y.~M.}\ \bibnamefont {Goh}},
  \bibinfo {author} {\bibfnamefont {S.~H.}\ \bibnamefont {Sung}}, \bibinfo
  {author} {\bibfnamefont {G.}~\bibnamefont {Ye}}, \bibinfo {author}
  {\bibfnamefont {S.}~\bibnamefont {Biswas}}, \bibinfo {author} {\bibfnamefont
  {D.~A.~S.}\ \bibnamefont {Kaib}}, \bibinfo {author} {\bibfnamefont
  {R.}~\bibnamefont {Dhakal}}, \bibinfo {author} {\bibfnamefont
  {S.}~\bibnamefont {Yan}}, \bibinfo {author} {\bibfnamefont {C.}~\bibnamefont
  {Li}}, \bibinfo {author} {\bibfnamefont {S.}~\bibnamefont {Jiang}}, \bibinfo
  {author} {\bibfnamefont {F.}~\bibnamefont {Chen}}, \bibinfo {author}
  {\bibfnamefont {H.}~\bibnamefont {Lei}}, \bibinfo {author} {\bibfnamefont
  {R.}~\bibnamefont {He}}, \bibinfo {author} {\bibfnamefont {R.}~\bibnamefont
  {Valentí}}, \bibinfo {author} {\bibfnamefont {S.~M.}\ \bibnamefont
  {Winter}}, \bibinfo {author} {\bibfnamefont {R.}~\bibnamefont {Hovden}},\
  and\ \bibinfo {author} {\bibfnamefont {A.~W.}\ \bibnamefont {Tsen}},\
  }\bibfield  {title} {\bibinfo {title} {Magnetic anisotropy reversal driven by
  structural symmetry-breaking in monolayer $\alpha$-{RuCl3}},\ }\href
  {https://doi.org/10.1038/s41563-022-01401-3} {\bibfield  {journal} {\bibinfo
  {journal} {Nature Materials}\ }\textbf {\bibinfo {volume} {22}},\ \bibinfo
  {pages} {50} (\bibinfo {year} {2023})}\BibitemShut {NoStop}%
\end{thebibliography}%

\end{document}